\documentclass[a4paper,11pt]{emulateapj}
\usepackage{amssymb, latexsym}
\usepackage{graphicx}
\usepackage{longtable}
\usepackage{amsmath}
\begin{document}
\title{Evolution of the Quasar Luminosity Function over $3 < \lowercase{z} <5$ in the COSMOS Survey Field}
\author{D. Masters\altaffilmark{1,2}, P. Capak\altaffilmark{3,4}, M. Salvato\altaffilmark{5,12}, 
	F. Civano\altaffilmark{6}, B. Mobasher\altaffilmark{1}, B. Siana\altaffilmark{1},  G. Hasinger\altaffilmark{7}, C. D. Impey\altaffilmark{8,9}, T. Nagao\altaffilmark{10},  J.R. Trump\altaffilmark{11}, H. Ikeda\altaffilmark{10}, M. Elvis\altaffilmark{11}, N. Scoville\altaffilmark{3}}

\altaffiltext{1}{Department of Physics and Astronomy, University of California, Riverside, CA 92521}
\altaffiltext{2}{Observatories of the Carnegie Institution of
  Washington, Pasadena, CA, 91101}
\altaffiltext{3}{California Institute of Technology, Pasadena, CA 91125}
\altaffiltext{4}{Spitzer Science Center, 314-6 Caltech, Pasadena, CA 91125}
\altaffiltext{5}{Max Planck-Institut f\"ur Extraterrestrische Physik,
Giessenbachstrasse 1, D-85748 Garching, Germany}
\altaffiltext{6}{Harvard-Smithsonian Center for Astrophysics, 60
Garden Street, Cambridge, MA 02138 USA}
\altaffiltext{7}{Institute for Astronomy, 2680 Woodlawn Drive, University of Hawaii, Honolulu, HI 96822}
\altaffiltext{8}{Department of Astrophysical Science, University of Princeton, Peyton Hall 103, Princeton NJ 08544, USA}
\altaffiltext{9}{Steward Observatory, University of Arizona, 933 North Cherry Avenue, Tucson AZ 85721, USA}
\altaffiltext{10}{Research Center for Space and Cosmic Evolution,
Ehime University, 2-5 Bunkyo-cho, Matsuyama 790-8577, Japan}
\altaffiltext{11}{University of California Obervatories/Lick Observatory and
Department of Astronomy and Astrophysics, University of California,
Santa Cruz, CA 95064 USA}
\altaffiltext{12}{Excellence Cluster Universe,
Giessenbachstrasse 1, D-85748 Garching, Germany}

\begin{abstract}

We investigate the high-redshift quasar luminosity function (QLF) down to an apparent magnitude of $I_{AB} = 25$ in the Cosmic Evolution Survey (COSMOS).  Careful analysis of the extensive COSMOS photometry and imaging data allows us to identify and remove stellar and low-redshift contaminants, enabling a selection that is nearly complete for type-1 quasars at the redshifts of interest.  We find 155 likely quasars at $z>3.1$, 39 of which have prior spectroscopic confirmation. We present our sample in detail and use these confirmed and likely quasars to compute the rest-frame UV QLF in the redshift bins $3.1 < z < 3.5$ and $3.5 < z < 5$. The space density of faint quasars decreases by roughly a factor of four from $z\sim3.2$ to $z\sim4$, with faint-end slopes of $\beta\sim-1.7$ at both redshifts. The decline in space density of faint optical quasars at $z > 3$ is similar to what has been found for more luminous optical and X-ray quasars. We compare the rest-frame UV luminosity functions found here with the X-ray luminosity function at $z > 3$, and find that they evolve similarly between $z\sim3.2$ and $z\sim4$; however, the different normalizations imply that roughly 75\% of X-ray bright active galactic nuclei (AGN) at $z\sim3-4$ are optically obscured. This fraction is higher than found at lower redshift and may imply that the obscured, type-2 fraction continues to increase with redshift at least to $z\sim4$. Finally, the implications of the results derived here for the contribution of quasars to cosmic reionization are discussed.  



\end{abstract}
\maketitle

\section{Introduction}

The evolution of the quasar luminosity function with redshift is a key observational constraint on the growth of supermassive black holes (SMBHs) over cosmic time \citep{Richstone98, Kaufmann00, Wyithe03, Marconi04}. The behavior of the QLF places constraints on the duty cycles of quasars, the growth history of SMBHs, and the coevolution of black holes and their host galaxies (\citealp{Hopkins06}, \citealp{Ueda03} and references therein). 

The QLF also determines the cumulative ionizing background radiation due to quasars. The faint end of the QLF at high redshift is of particular interest in this regard because faint quasars contribute substantially to the total ionizing background due to quasars. While the peak in quasar activity around $z\sim2-3$ is responsible for HeII reionization at $z\sim3$ \citep{Reimers97, Sokasian02}, the relative contribution of quasars to hydrogen reionization at $z\sim6-10$ is not as well constrained. The observed decline in space density of highly luminous quasars at $z > 3$ has been taken as evidence that quasars contribute negligibly to hydrogen reionization \citep{Madau99, Fan06}. However, only the most luminous quasars at high redshift can be found in surveys such as SDSS, leaving the contribution of faint quasars to the ionizing background unknown. 

The faint end is challenging to study at high redshift because of the need for survey fields that are both deep enough to detect faint sources and wide enough to find a statistically significant number of quasars. In addition, follow-up spectroscopy is difficult due to the faintness of the sources. Two groups have recently investigated the faint end of the QLF at $z\sim4$: \citealp{Glikman11} (hereafter G11) using parts of the Deep Lens Survey (DLS, \citealp{Wittman02}) and NOAO Deep Wide-Field Survey (NDWFS, \citealp{Jannuzi99}) fields, with a combined sky coverage of $3.76~\mathrm{deg}^{2}$, and \citealp{Ikeda11} (hereafter I11) using the HST-ACS region of the COSMOS field \citep{Scoville07} with a sky coverage of $1.64~\mathrm{deg}^{2}$. The value of the faint-end slope $\beta$ measured in G11, $-1.6^{+0.8}_{-0.6}$, is consistent with the value of $-1.67^{+0.11}_{-0.17}$ reported in I11 for the COSMOS field. 

While these studies agree on the faint-end slope of the QLF at $z\sim4$, they disagree on the absolute space density of low-luminosity quasars by roughly a factor of four, with a higher space density reported in G11. This discrepancy cannot be attributed to cosmic variance (see \S~8.1), and leads to different pictures of quasar evolution. The result reported in G11 implies that the decline in the space density of faint quasars with redshift after the peak at $z\sim1-2$ eventually stops and possibly reverses, which could make the contribution of quasars to cosmic reionization significant.

Both I11 and G11 use broad-band optical color selection to identify quasars. A significant uncertainty in this approach is the selection function, or the fraction of quasars that are selected as a function of redshift and magnitude. The selection function is usually determined with Monte Carlo simulations, which are sensitive to the assumed distribution of quasar spectral energy distributions (SEDs) in the survey field. This distribution is uncertain, and mismatches between the assumed and actual distributions can give rise to significant errors in completeness estimation and thus the derived luminosity function. 

In addition to selection function uncertainty, contamination from stars and galaxies can be substantial and difficult to quantify. Spectroscopy allows a robust determination of the contamination, but due to the large amount of telescope time required, often only a subset of the candidates can be observed. Extrapolating to the complete sample can result in large errors in the estimated contaminant fraction.

Here we take a different approach (described in \S~2), using the COSMOS survey data to select and identify high-redshift quasars with high completeness and low contamination.  Our principal goals are to (1) determine the faint end of the QLF from $3.1 < z <5$ in COSMOS, thereby providing an independent check of the result reported by I11, as well as a direct determination of the evolution of the faint-end over this redshift range, (2) compare the rest-frame UV quasar luminosity function with the rest-frame 2-10 keV X-ray luminosity function for sources at $z>3$ in order to investigate the obscured fraction of bright AGN at these redshifts, and (3) use our results to make inferences regarding the evolution of the QLF and the likely contribution of quasars to reionization, both of HeII at $z\sim3$ and HI at $z\sim6-10$. 

The structure of this paper is as follows. In \S~2 we provide an overview of the method used to determine the QLF. In \S~3 we present our initial quasar selection. In \S~4 we describe the tests used to verify the high completeness of the selection. In \S~5 we describe how we separate stellar and low-z contaminants from the high-redshift quasar population of interest. In \S~6 we present the final sample of confirmed and likely quasars and discuss the X-ray properties of the sample. In \S~7 we use the final list of likely quasars to compute the luminosity function at $z\sim3.2$ and $z\sim4$. In \S~8 we discuss sources of error that may influence our results. In \S~9 we compare our results with the X-ray luminosity function at similar redshifts and discuss the implications for the obscured quasar fraction at these redshifts. In \S~10 we discuss the ionizing background due to quasars. In \S~11 we present our main conclusions. 

We assume a cosmology with $\Omega_{\Lambda} = 0.7$, $\Omega_{M} = 0.3$, and $\mathrm{H}_{0} = 70~ \mathrm{km}~\mathrm{s}^{-1}~\mathrm{Mpc}^{-1}$. Magnitudes are in the AB system.

\section{Overview of Methodology}

\begin{figure*}
        \centering
	\includegraphics[scale=0.55]{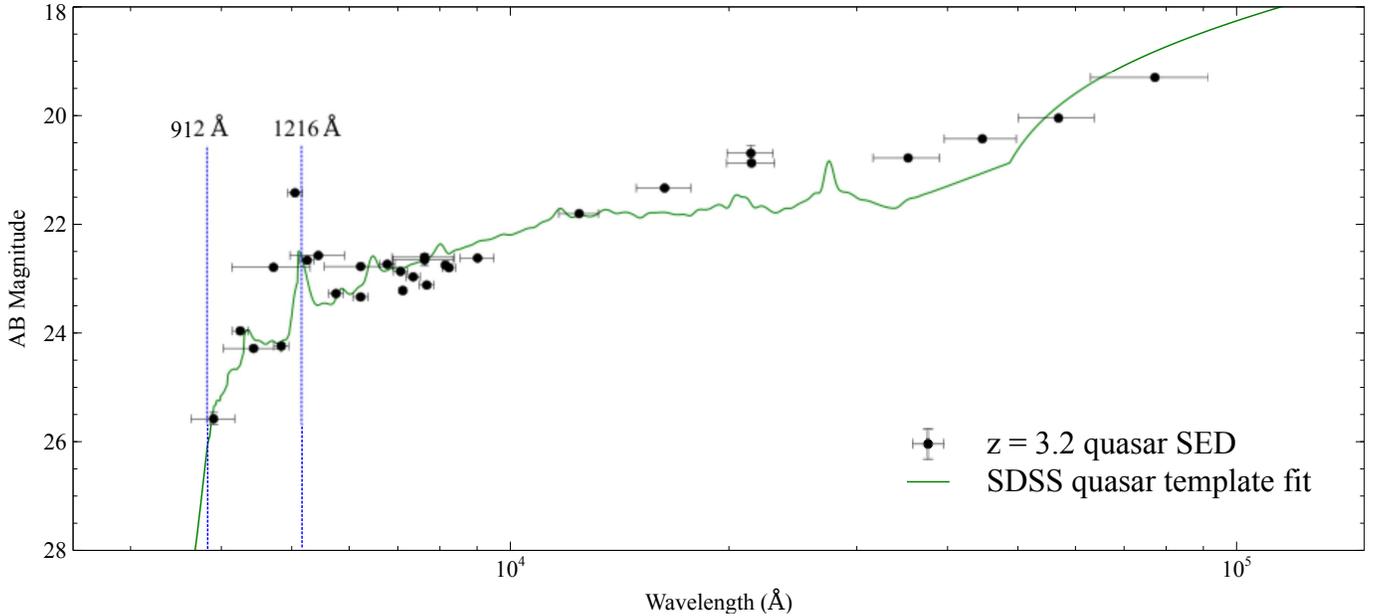}
	\caption{This confirmed quasar at redshift $z=3.2$ illustrates the quality of the low-resolution spectrum afforded by the COSMOS photometric data. Filter bandwidth is indicated with horizontal bars. The observed-frame positions of the Lyman break and Lyman limit are indicated with vertical blue lines. Overlaid is a typical QSO spectral template \citep{Vanden01}. Features of a high-redshift quasar SED are clear in the photometry, including the Lyman break at rest-frame 1216\AA, Ly$\alpha$ emission, broad quasar emission lines, and the rising SED into the observed-frame infrared.}
\label{Fi:sed_example}
\end{figure*}

We utilize the COSMOS broad, intermediate, and narrow band photometric catalog \citep{Capak07, Sanders07} for the selection of quasars and identification of contaminants. With 29 bands of well-matched photometry, these data constitute low-resolution spectra for all objects over the wavelength range 0.1-8$\mu$m (Figure \ref{Fi:sed_example}). This is generally sufficient to distinguish stars from high-redshift quasars. With spectra of sufficient quality for all sources in the field, the population of quasars can be well-constrained down to the limiting magnitude. Our approach is similar to that of \citet{Wolf03}, who used the 17 filters (5 broad and 12 intermediate band) of the COMBO-17 survey to identify quasars. COSMOS is of similar resolution to the COMBO-17 survey in the optical, but significantly deeper and covering a wider wavelength range. 

Because neither the properties of the faint, high-redshift quasar population nor the properties of the contaminating stellar population are very well constrained, we individually inspect the spectral energy distributions and imaging data to identify quasars and reject contaminants, as described in \S~5. To ensure that our classifications are unbiased, we also cross-check against automated $\chi^{2}$ fitting of model SEDs to the low resolution spectro-photometry of COSMOS (\S~6.2).  The two methods produce similar results, but we argue the visual classification is more accurate because it allows for a broader range of quasar properties as well as the identification of photometry errors that can influence the $\chi^{2}$ result. 

The primary method we use can be summarized as follows:

\begin{enumerate}
\item{Find a coarse selection (based on point-source morphology, the presence of a Lyman-break, and a power law infrared slope) that is highly complete for type-1 quasars at $z\sim3-5$, sacrificing reliability to the extent necessary to achieve high completeness.}
\item{Assess the 29-band photometric and imaging data for each candidate to remove contaminants.}
\item{Cross-check the resulting quasar list with known high-redshift quasars in COSMOS to confirm our ability to recover quasars with high completeness.}
\item{Compute the luminosity function with the resulting sample.}
\end{enumerate}

Several factors specific to the COSMOS field make this approach reasonable. We list these below.

\begin{itemize}
\item{Over 40 confirmed $z > 3$ quasars are known in COSMOS, which were selected in different ways (X-ray, infrared, optical). These quasars serve as a guide in developing our selection criteria, and provide an important check on our ability to distinguish quasars from contaminants using photometric and imaging data.} 
\item{A large spectroscopic sample of faint AGN and galaxy candidates obtained by Keck and VLT can be used to test contamination.}  
\item{High-resolution imaging of the COSMOS field with the HST lets us restrict our sample to true point sources, limiting the contamination from high-redshift star-forming galaxies.}
\item{Accurate photometric redshifts \citep{Ilbert09, Salvato09} can be used for sources lacking spectroscopic confirmation.}
\item{Deep Chanda X-ray imaging of the central $0.9~\mathrm{deg}^{2}$ of the COSMOS field allows another check on the completeness and reliability of our approach, and also permits a comparison of our estimated QLF with the X-ray luminosity function at similar redshifts. We present this analysis in \S~9.}
\end{itemize}

\section{Initial Candidate Selection}

Spectroscopic campaigns, e.g. \citet{Lilly07}, \citet{Trump09}, \citet{Ikeda11}, and Capak et al. (in preparation), have confirmed over 40 quasars at $z > 3$ in the COSMOS field. These known quasars were originally selected in different ways (X-ray, infrared, optical) and therefore display a range of SEDs. These confirmed sources are used to guide the development of a coarse initial selection, which is then tested for completeness with simulated quasar photometry, as described in \S~4. 

It is worth emphasizing that the selection presented here is intended only to be a weak filter against contamination, in order to reduce the number of candidates to a reasonable number while maintaining high completeness. The careful rejection of contaminants occurs later through inspection of the photometric and imaging data, as described in \S~5.


We begin by selecting point sources in the magnitude range $16 \leq I \leq 25$, restricting our study to the region of COSMOS covered by ACS imaging (1.64~$\mathrm{deg}^{2}$). Spatially extended sources as defined in the Hubble Advanced Camera for Surveys (ACS) catalog of \citet{Alexie07} are excluded. The test for point morphology is based on the ratio of the peak surface brightness (MU\_MAX) to the magnitude (MAG\_AUTO), and exploits the fact that the light distribution of a point source scales with its magnitude. The point source selection is shown to be robust to $I=25$, which sets the magnitude limit for our search. 22383 candidates meet this selection criterion.
   
\begin{figure}[th]
       \centering
	\includegraphics[scale=0.48]{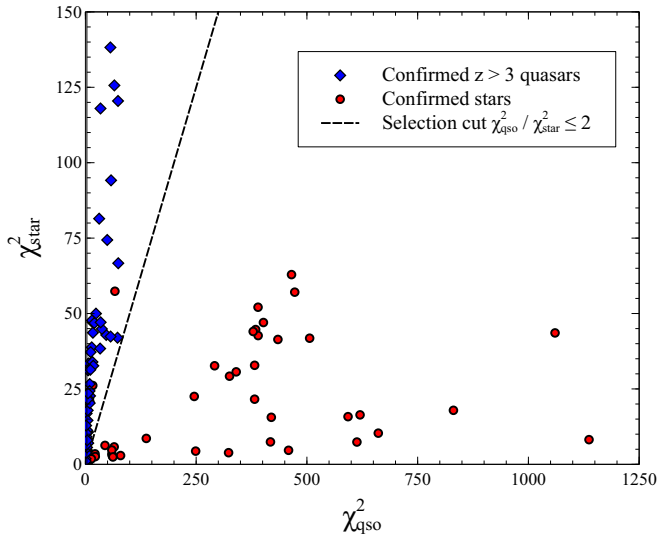}
	\caption{Comparison of the minimum $\chi^{2}$ values for QSO and star template fitting to the COSMOS photometry of point sources with spectra. The cut shown selects objects with $\chi^{2}_{qso} / \chi^{2}_{star} \leq 2.0$, which retains all known $z>3$ quasars while rejecting a significant fraction of stars. The spectroscopically confirmed stars shown here are primarily F, M, L, and T dwarfs, which are typical contaminants in high-redshift quasar searches (\citealp{Richards09}).}
\label{Fi:plotone}
\end{figure}

Next we use the $\chi^{2}$ parameter, computed through template fitting in the COSMOS photometric redshift catalog, to separate stars from quasars.  The $Le~Phare$ code \citep{Arnouts11} was used to determine photometric redshifts (see  \citealp{Ilbert09} for details). As a result of extensive fitting against templates, sources are assigned three $\chi^{2}$ values: $\chi^{2}_{gal}$, $\chi^{2}_{qso}$, and $\chi^{2}_{star}$, which are computed by fitting galaxy, AGN, and stellar templates, respectively, to the source photometry. Because we restrict our search to point sources, a comparison between $\chi^{2}_{qso}$ and $\chi^{2}_{star}$ can distinguish most stars from quasars. As seen in Figure \ref{Fi:plotone}, the relaxed cut $\chi^{2}_{qso} / \chi^{2}_{star} \leq 2.0$ selects all known $z > 3$ quasars, while rejecting the majority of stars. This leaves us with 8125 candidates.

The effectiveness of $\chi^{2}$ fitting is due to two main factors. One is the COSMOS intermediate band data, which can differentiate sharp features such as the Lyman break and emission features in quasar photometry from the smooth photometry of stars. The other (more important) factor is Spitzer IRAC photometry. Stars display Rayleigh-Jeans black body curves that decline in the infrared, whereas high redshift quasars display power law SEDs that are flat or rising in the infrared.

Next we apply a cut designed to select high-redshift quasars. The Lyman break, caused by the scattering of radiation at wavelengths shorter than 1216~\AA\ by neutral hydrogen in the IGM and quasar host galaxy, results in a reduction in $U$ band flux for quasars at $z \gtrsim 3.1$. In the mid-infrared, however, high redshift quasars remain relatively bright due to their intrinsic power law spectral slopes. Therefore, we accept sources meeting either of the following cuts: $U - \mathrm{ch}1 \geq 1.5$ or $U - \mathrm{ch}2 \geq 1.5$. Here ch1 and ch2 refer to Spitzer IRAC channel 1 (3.6$\mu$m) and channel 2 (4.5$\mu$m). These cuts were motivated by the spectroscopically confirmed sample as well as simulated quasar spectra, and are significantly relaxed in order to achieve high completeness. While the cuts are not significantly different, we accept sources meeting either one in order to avoid losing quasars with bad IRAC photometry in one of the bands. This leaves 6667 candidates. To safeguard against losing unusual quasars, we also accept sources with photometric redshifts $z_{qso} \geq 3$. This adds 28 candidates, for a total of 6695 candidates after this step. 

Finally, we require that the source is detected at a significant level ($m_{AB} \leq 24.0$) in either IRAC channel 1 or channel 2. Because quasar SEDs typically rise into the infrared, this condition will be met by the majority of quasars with $I < 25$, while it serves to eliminate faint stars or other unexpected contaminants. The final number of candidates after applying this selection is 4009. 

This selection, by utilizing only the Lyman continuum break and power law infrared emission of high-redshift quasars, largely avoids the uncertainties associated with broad-band color selection. Color selections in the rest-frame optical must correctly account for intrinsic quasar SEDs that can vary substantially in spectral slope and broad line emission equivalent widths, the distributions of which are not well constrained at high redshift. In addition, this selection should be sensitive to broad absorption line (BAL) quasars, which show similar mid-infrared to optical luminosity ratios as non-BALs \citep{Gallagher07}.

To summarize, the initial (coarse) selection we use is:

\begin{enumerate}
\item{16 $< I(\mathrm{mag\_auto}) <$ 25}
\item{Point source based on ACS $I$-band imaging}
\item{$\chi^{2}_{qso} / \chi^{2}_{star}$ $\leq 2.0$}
\item{$U - \mathrm{ch}1 \geq 1.5$ OR $U - \mathrm{ch}2 \geq 1.5$ OR $z_{qso} \geq 3.0$ }
\item{$m_{AB}(3.6\mu\mathrm{m}) \leq 24.0$ OR $m_{AB}(4.5\mu\mathrm{m}) \leq 24.0$}
\end{enumerate}

The 4009 candidates selected include all 39 known $z \geq 3.1$ quasars in COSMOS. 

A comparison of our selection with the high-redshift X-ray sources reported in \citet{Civano11} confirms that we select all high redshift X-ray sources with $I \leq 25$ and point morphology (27 total). The other 74 sources in the X-ray $z > 3$ sample are primarily classified as extended (90\%), or lie outside our $I$-band magnitude limit, or both. Keck DEIMOS spectra (Capak et al., in preparation) of 12 of the extended $z > 3$ X-ray sources confirms that these are obscured, type-2 AGN without broad emission lines, and therefore not appropriate to include in this work. 

These pieces of evidence indicate that the selection outlined above is highly complete for unobscured quasars at $z > 3.1$. This is verified using simulated quasar photometry, as described in the next section.

\section{Selection Completeness}

We follow the standard procedure of testing the selection by simulating a large number of quasars in redshift/magnitude space. We use the three composite spectral templates QSO1, TQSO1, and BQSO1 from the SWIRE template library \citep{Polletta07}, each representing a different infrared/optical flux ratio, and extend them into the far-UV using the composite HST spectral template of \citet{Telfer02}. By letting the intrinsic spectral slopes for $\lambda >$ 1216~\AA\ and $\lambda <$ 1216~\AA\ vary randomly according to the distributions given in \citet{Vanden01} and \citet{Telfer02}, we generate 50 QSO1 templates and 25 each of the TQSO1 and BQSO1 templates. These 100 base templates are checked against the photometry of known quasars in SDSS, including IRAC photometry for quasars in the Spitzer Heritage Archive \citep{Teplitz12}, to verify that they are representative of the underlying variation in the population of type-1 quasars.

We create a grid in redshift and $I$ magnitude, with the redshift extending from $3 \leq z \leq 5$ in steps of $\Delta z = 0.1$ and the $I$ band magnitude extending from $20 \leq I \leq 26$ in steps of $\Delta I$ = 0.1. At each point in this grid we generate 20 realizations of each of the 100 model quasar templates, adding Poisson and background noise to the measured flux in each filter, and correcting for intergalactic extinction using the model of \citet{Madau95}. The selection criteria (4) and (5) are then applied to each of the 2000 resulting SEDs to determine the completeness at that grid point. 

\begin{figure}[ht]
       \centering
	\includegraphics[scale=0.5]{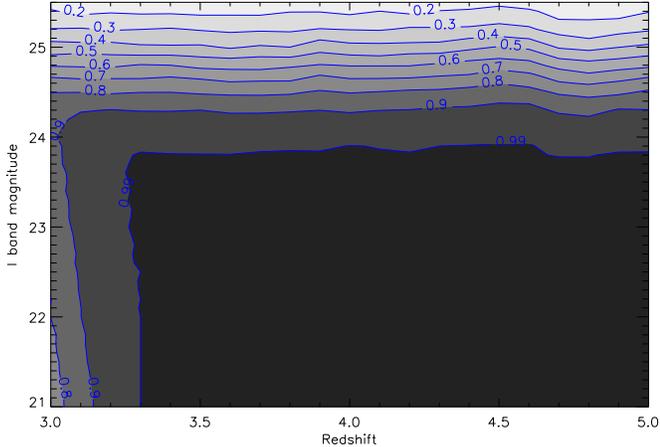}
	\caption{Completeness map for our photometric selection. For magnitudes brighter than $I\sim24$, the selection is over 90\% complete over the redshift range $3.1 \lesssim z \lesssim 3.3$ and $\sim$100\% complete at higher redshifts. Because we also accept all sources with $z_{qso} \ge 3$, the completeness from $3.1 \lesssim z \lesssim 3.3$ will be enhanced over what is shown. At $I$-band magnitudes fainter than $I\sim24$ there is some incompleteness due to criterion (5), which is accounted for in computing the luminosity function.}
\label{Fi:completeness3}
\end{figure}

The result is shown in Figure \ref{Fi:completeness3}. For $I < 23.8$, the completeness is $>$90\% over the redshift range $3.1 < z < 3.3$ and $\sim$100\% at higher redshifts. The marginal incompleteness for $3.1 \lesssim z \lesssim 3.3$ will be improved by the additional inclusion of sources with $z_{qso} > 3$ in criterion~(4). There is some incompleteness for $23.8 < I < 25$ due to criterion~(5), which is accounted for when computing the luminosity function. 

Figure \ref{Fi:completeness3} demonstrates that the photometric criteria (4) and (5) are highly complete, but it does not account for possible incompleteness introduced by criteria (2) and (3). We briefly discuss the effect these may have.  

\begin{figure*}
        \centering
	\includegraphics[scale=0.55]{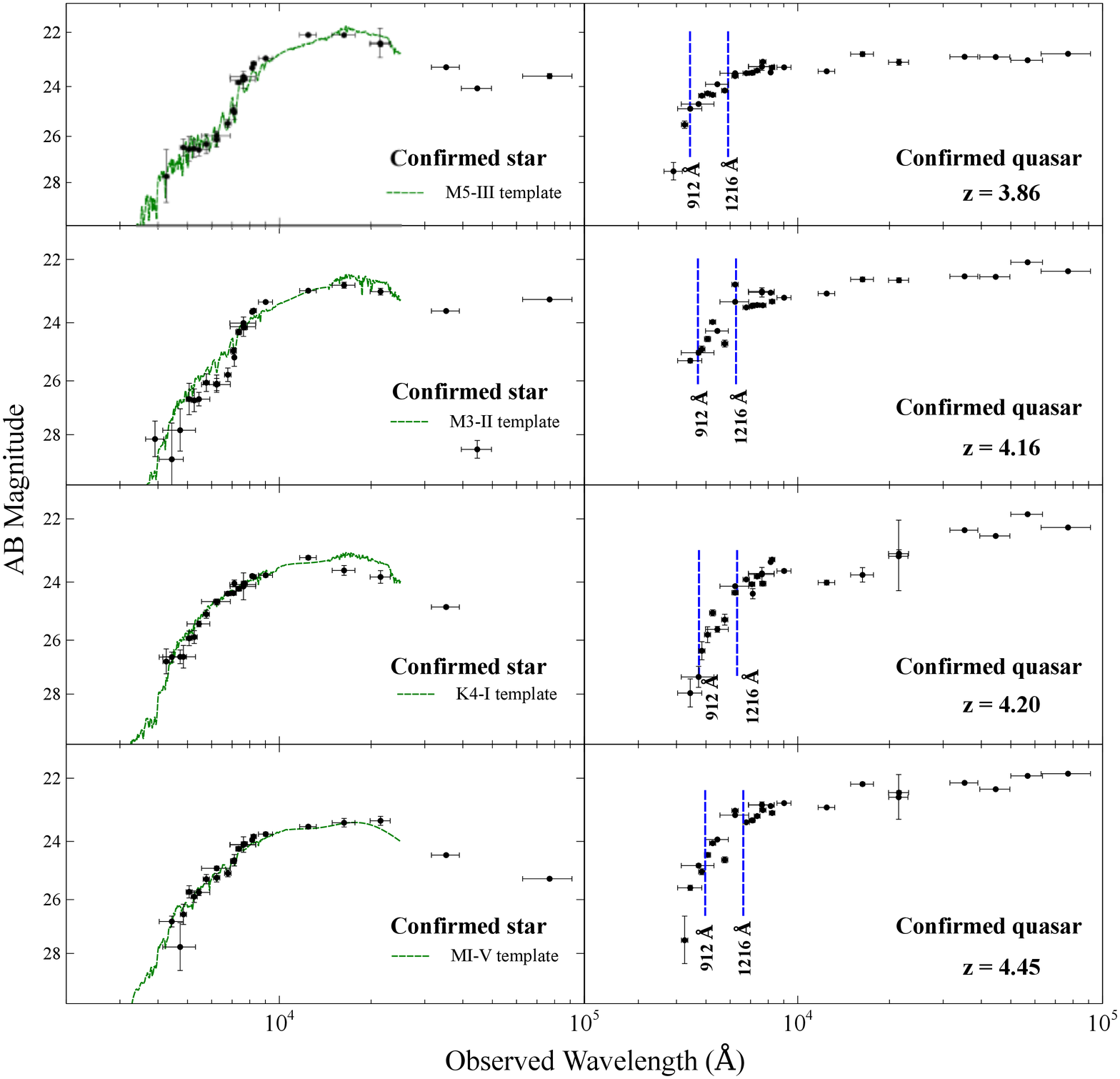}
	\caption{SEDS from the quasar candidates selected in I11. Stars are shown on the left and confirmed quasars on the right. The stars are dwarfs representative of those typically contaminating quasar searches, and the quasars are among the faintest known. Stellar templates \citep{Pickles98} are overlaid in green. The obvious differences between the SEDs of the objects make it relatively simple to distinguish them, even without high-resolution spectroscopy. The observed-frame position of the the Lyman break and Lyman limit are indicated on the quasar SEDs with vertical blue lines. The distinct break, with a power law SED in the infrared bands, is characteristic of faint, high redshift quasars and distinguishes them from stars.}
\label{Fi:plottwo}
\end{figure*}

The third selection criterion ($\chi^{2}_{qso} / \chi^{2}_{star}$ $\leq 2.0$) is based on the results of extensive fitting of the COSMOS sources against spectral templates. Checking this criterion with our modeled quasar templates would not be very revealing, as it would constitute checking simulated photometry against simulated photometry. However, the evidence in Figure \ref{Fi:plotone} gives us confidence that this very relaxed criterion introduces negligible incompleteness.

The point source criterion can potentially induce incompleteness because ACS imaging, with a point-spread function (PSF) with FWHM of $0.12''$ at the detector, resolves all but the most compact of objects. While this is desirable in that it excludes galaxies from consideration, it may also reject intrinsically faint quasars with host galaxy light contributing to the overall flux. We discuss this issue in more detail in \S~8.4. 

\section{Examination of Candidates and Identification of Quasars}

The selection described above is highly complete but also unreliable, in the sense that a small fraction of the selected objects actually are quasars at the redshifts of interest. However, careful examination of the COSMOS photometric and imaging data for each of the 4009 candidates is relatively fast and allows us to confidently reject typical contaminants such as dwarf stars and lower-redshift quasars.


Visual examination of the data allows more robust classification than a reliance on SED template fitting. The 29 bands of photometry and corresponding imaging data reveal stars with unusual colors that mimic quasar SEDs closely, often because of photometric contamination, as well as true quasars that automated tests fail to find because of contaminated or unusual photometry. In addition, coarse redshift estimates can be made based on the position of Ly$\alpha$ emission and/or the Lyman break, which can then be compared with the results from automated fitting.

The relevant high redshift quasar characteristics we look for in the COSMOS photometric and imaging data include: (1) a strong spectral break at rest-frame 1216~\AA, (2) a power law SED that is flat or rising in the infrared IRAC bands, (3) Ly$\alpha$ emission evident in intermediate band filters, and (4) often other strong rest-frame UV emission lines (e.g. CIV~$\lambda$1549, CIII]~$\lambda$1909) evident in intermediate band filters as well. 

Stars, in contrast, have SEDs with characteristic Rayleigh-Jeans shapes that decline in the IRAC bands. The decline in the blue side of a stellar SED is less abrupt than the Lyman break in the SED of a high redshift quasar, and there are no emission lines in the intermediate bands.  Figure \ref{Fi:plottwo} illustrates, for some of the most difficult cases, the clear differences in the SEDs of high redshift quasars and stellar contaminants.

The candidates were examined using Specpro \citep{Masters11}, an interactive IDL tool designed primarily for spectral analysis in the context of multiwavelength surveys\footnote[1]{specpro.caltech.edu}. This program displays the imaging and photometric data for each candidate and allows us to perform rough SED fitting and redshift estimation. The first author went through all candidates after significant calibration against the SEDs of known stars and high-redshift quasars, and subsets of the data were examined by coauthors to establish the consistency and reproducibility of classification. It was found that classifications were highly consistent among different classifiers.

As expected, our sample includes a large fraction of stellar contaminants. Most of these are dwarf stars, while a fraction are stars whose optical fluxes are incorrectly matched to a nearby IR-bright source, which is apparent through examination of the stamp images. The mismatch artificially boosts the SED in the IRAC bands, such that the $\chi^{2}_{qso}$ vs. $\chi^{2}_{star}$ comparison does not favor a star assignment. 

We also find a number of quasars or compact galaxies at $z < 3.1$ in our sample, which reflects the fact that our Lyman-break criterion (4) is very relaxed. This can also be seen in the slow decline in completeness toward $z=3$ in Figure \ref{Fi:completeness3}.  Lower-redshift objects are identified by the position (or absence) of the Lyman break / Ly$\alpha$ emission in the SED. Of the contaminants, we estimate that 28\% are $z < 3.1$ objects (quasars or compact low-$z$ galaxies), and 64\% are stars. The remaining 8\% are unclear based on the photometry but likely fall in one of these categories as they do not show characteristics of $z>3.1$ quasar SEDs. 

For the sources identified as likely high redshift quasars, we assign a confidence flag ranging from 2 to 4, where 4 is highly confident and 2 is somewhat confident. We emphasize that sources even marginally consistent with being high-redshift quasars are retained (usually at confidence 2) in order to maintain high completeness. Candidates assigned a confidence of 4 typically are X-ray detected and/or show strong emission lines in intermediate bands, as well as a clearly defined Lyman break and rising infrared SED. Candidates assigned a confidence 3 show a well-defined Lyman break and/or Ly$\alpha$ emission evident in an intermediate band filter as well as a rising infrared SED, but are not clearly detected in X-ray. We are confident that these are high redshift quasars based on their point-source morphology and SED. Candidates assigned a confidence 2 show a moderately convincing photometric break and are flat or rising in the infrared, but do not necessarily show emission features in intermediate bands, nor are they X-ray detected. Of our candidates these are the most likely to be either very compact galaxies or unusual stars rather than quasars. Follow-up spectroscopy on these sources may be required to refine the results presented here.

\section{Quasar Sample}

Of 4009 candidates, we find 155 likely type-1 quasars at $z > 3.1$. Of these, 48 are considered confidence 2 sources, 54 are considered confidence 3, and 53 are considered confidence 4. The final list of the 155 confirmed and likely $z > 3.1$ quasars is given in Table \ref{Ta:quasars} in the Appendix. 

\subsection{Identification Reliability}
\begin{figure}[th]
       \centering
	\includegraphics[scale=0.4]{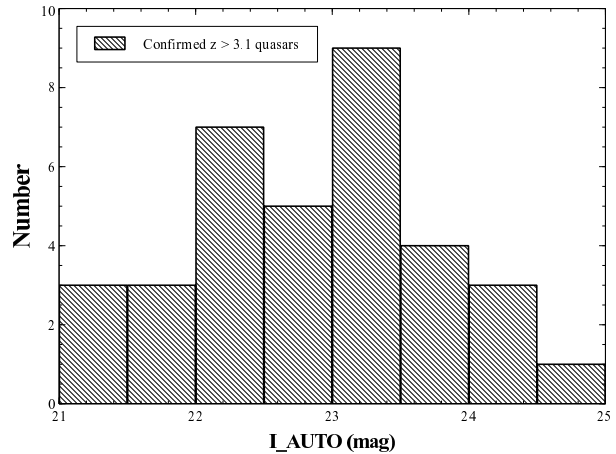}
	\caption{The distribution in  $I(\mathrm{mag\_auto})$ for the spectroscopically confirmed subset of our final quasar candidates. We have recovered both relatively bright and very faint known quasars in the field, indicating that there is high overall completeness in our classification method regardless of quasar luminosity.}
\label{Fi:imags}
\end{figure}

An important point is that the quasar candidates were examined without knowing which were the previously confirmed quasars, yet every one of the 39 known quasars at $z \geq 3.1$ in the COSMOS field was correctly classified based on the imaging and photometric data. The method employed thus independently recovered every previously known type-1 quasar in the field. Figure \ref{Fi:imags} shows that the confirmed sample includes a number of very faint sources, including those discovered by \citet{Ikeda11}. This is strong evidence that we do not miss actual high-redshift quasars in the initial sample. If anything, we may overcount by misidentifying either stars or high-redshift galaxies as quasars. Because stars display easily recognizable SEDs, we consider it unlikely that they constitute a significant contaminant. 

While stars are relatively easy to reject, high redshift galaxies display a similar overall SED to a quasar and could contaminate the sample. We expect that few of the candidates identified as quasars are actually compact galaxies, because the strict point-source criterion will eliminate nearly all galaxies at these redshifts. We discuss this potential source of error in more detail in \S~8.3.

\subsection{Comparison with Automated $\chi^{2}$ Selection}
As another check on the reliability of our classifications, we compare our final sample with a purely automated selection using the results of the extensive $\chi^{2}$ fitting against stellar and AGN templates described in \citet{Ilbert09}. Of the initial sample of 4009 candidates, 195 objects have photometric redshift estimates above $z=3.1$. Limiting the sample to objects with $\chi^{2}_{qso} / \chi^{2}_{star} \leq 1.0$ (i.e., quasar is the favored solution from $\chi^{2}$ fitting) the final automated sample is 143 objects, in very close agreement with our final sample of 155 from visual classification. 

\begin{figure}[th]
       \centering
	\includegraphics[scale=0.4]{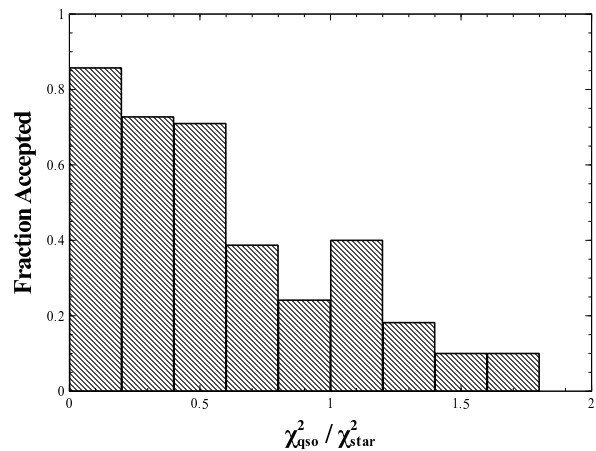}
	\caption{The fraction of the $z_{qso} > 3.1$ candidates we identify as high redshift quasars as a function of the ratio $\chi^{2}_{qso}/\chi^{2}_{star}$. As expected, the lower this ratio, the more likely that we identify a candidate as a quasar through inspection of the imaging and photometry, showing that our visual inspection is in accord with the results of $\chi^{2}$ fitting against spectral templates.}
\label{Fi:ratiotest}
\end{figure}

Figure~\ref{Fi:ratiotest} shows that, of the 195 objects with photometric redshifts above 3.1, our visual classification typically retains objects with $\chi^{2}_{qso}/\chi^{2}_{star} \lesssim 1$ and rejects objects with  $\chi^{2}_{qso}/\chi^{2}_{star} \gtrsim 1$, as would be expected if the visual classification is in agreement with $\chi^{2}$ fitting.

Of the 143 objects that would be selected automatically, 91 are in common with our final selection. The sources that we rejected but automated fitting would retain are, for the most part, stars mismatched to a nearby IR-bright source, artificially elevating their IRAC fluxes and thus lowering their $\chi^{2}_{qso}$ value. This is reflected in the fact that these sources have higher average IRAC-to-optical match distances.

On the other hand, we identify 64 likely quasars that are rejected by the automated selection. These primarily are rejected due to a low photometric redshift estimate. However, they are consistent with being high-z quasars based on a weak photometric break and power law infrared spectrum, so we retain them.

Despite the slightly different samples, it is clear that automated selection based on $\chi^{2}$ fitting produces essentially the same answer as careful rejection of contaminants on an object-by-object basis. 

\subsection{X-ray Properties}

The Chandra COSMOS survey (C-COSMOS, \citealp{Elvis09}) is a 1.8 Ms Chandra program covering the central  $0.5~\mathrm{deg}^{2}$ of the COSMOS field with an effective exposure of $\sim$160 ks, and an outer  $0.4~\mathrm{deg}^{2}$ area with an effective exposure of $\sim$80 ks. In addition, the entire COSMOS field has been observed in the X-ray with XMM-Newton \citep{Hasinger07} to a brighter X-ray flux limit.

Of the final sample of confirmed and likely quasars presented here, 27 are X-ray detected by Chandra (\citealp{Civano11}, 2012 submitted) and 12 are detected by XMM \citep{Brusa10}. The remainder of the objects are undetected and therefore have X-ray fluxes below the detection limits of either Chandra deep, Chandra shallow, or XMM, depending on their location in the COSMOS field. We perform a stacking analysis of 47 unconfirmed sources without X-ray detection in the Chandra region to investigate whether their X-ray fluxes are consistent with expected quasar emission.

\begin{figure}[ht]
       \centering
	\includegraphics[scale=0.42]{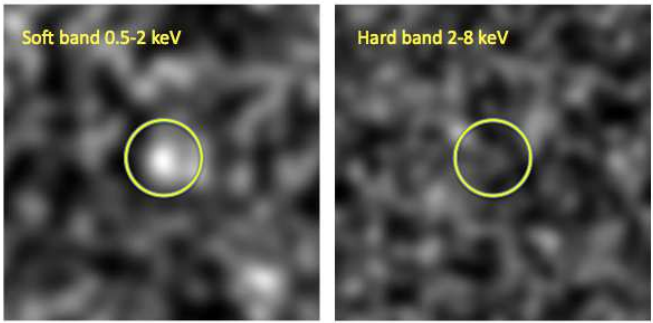}
	\caption{Stacking analysis of 47 likely quasars that fall in the Chandra region but are not individually detected in X-ray. The images are $30''\times30''$ and are smoothed with a Gaussian kernel with FWHM of $6''$. The $\sim$5-6 sigma detection in the soft band (corresponding to the hard X-ray band in the rest frame at $z\sim3-4$) is evidence that at least a fraction of the individually undetected sources are quasars. The lack of a clear detection in the observed-frame 2-8 keV stack indicates they are likely unobscured, intrinsically faint objects that fall below the detection limit in the hard band.}
\label{Fi:xray_stack}
\end{figure}

We use CSTACK \citep{Miyaji08} to compute the number of X-ray counts at the optical position of the undetected sources.
This program correctly accounts for the different exposure times depending on source position. We find a detection in the soft-band 0.5-2.0 keV (Figure \ref{Fi:xray_stack}), corresponding closely to rest-frame 2-10 keV for our sources, with a significance of $\sim$5-6 sigma, as determined through random stacking in blank fields. We find no clear detection in the observed-frame 2-8 keV hard band. The stacked detection in the soft band provides evidence that these sources are quasars with individual X-ray fluxes below the Chandra detection limit, while the lack of a clear hard band detection suggests no significant X-ray obscuration. 

\begin{figure}[ht]
       \centering
	\includegraphics[scale=0.5]{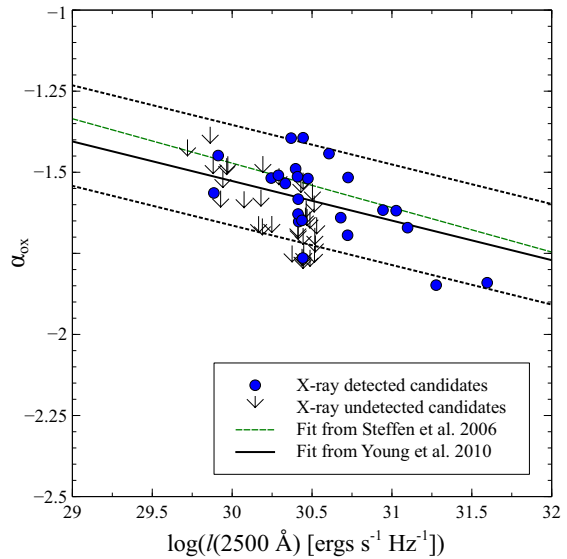}
	\caption{The X-ray / UV correlation for the sources in our sample in the Chandra region of COSMOS. The $\alpha_{ox}$ ratios (or upper limits on the ratio for sources without individual X-ray detection) are generally consistent with measured correlations at lower redshifts. The dotted lines indicate the 1-sigma dispersion in the $\alpha_{ox}$ relation of Young et al. 2010.}
\label{Fi:alpha}
\end{figure}

As another check on whether the X-ray flux limits on the nondetected candidates are consistent with these objects being quasars, we compute the $\alpha_{ox}$ ratio for sources in the C-COSMOS region, making use of the SEDs to derive the rest-frame UV fluxes. The $\alpha_{ox}$ parameter represents the spectral slope between the UV and X-ray flux$\colon$

\begin{equation}
 \alpha_{ox} \equiv \frac{\mathrm{log}(f_{\mathrm{X}}/f_{\mathrm{uv}})}{\mathrm{log}(\nu_{\mathrm{X}}/\nu_{\mathrm{uv}})}
\end{equation}

where $f_{\mathrm{uv}}$ is the 2500~\AA\ flux and  $f_{\mathrm{X}}$ is the 2 keV X-ray flux. For X-ray detected sources we derived the 2keV X-ray flux from the 0.5-2 keV flux \citep{Elvis09}. For sources without X-ray detection, we compute an upper limit on the X-ray flux, taking into account the X-ray flux limit at the optical position of each source. The result in Figure~\ref{Fi:alpha} shows that the nondetected sources tend to have lower UV luminosity, with upper limits on the $\alpha_{ox}$ ratio that are generally consistent with previously measured correlations between $\alpha_{ox}$ and the 2500~\AA\ luminosity (\citealp{Steffen06}, \citealp{Young10}). In conjunction with the stacking results, this gives additional evidence that the sources are faint, type-1 quasars with X-ray fluxes below the Chandra detection limit. 

\section{Luminosity Function}

We derive the luminosity function for the redshift bins $3.1 \leq z \leq 3.5$ and $3.5 < z \leq 5.0$, using the standard $1/V_{max}$ estimator \citep{Schmidt68}. 

The luminosity function for quasars is generally found to be well-described by a double power law of the form
\begin{equation}
\resizebox{.9\hsize}{!}{$\Phi(M,z) = \frac{\Phi(M^{*}) }{10^{0.4(\alpha+1)(M-M^{*})}+10^{0.4(\beta+1)(M-M^{*})}}$}
\end{equation} 
where $\Phi(M^{*})$ is the normalization, $M_{1450}^{*}$ is the break luminosity between the bright and faint ends, $\alpha$ is the bright-end slope and $\beta$ is the faint-end slope \citep{Boyle88}. We use this parameterization to fit our results in conjunction with bright-end results of \citet{Richards06} derived in the SDSS. Before presenting the results we first outline the determination of redshift and absolute magnitude for the sample.  

\subsection{Redshift Determination}

Only 39 of the 155 sources in the final sample have spectroscopic redshifts. For those that do not, we use the photometric redshift $z_{qso}$ determined by fitting AGN templates to the source SED \citep{Ilbert09}. In the cases in which a candidate is X-ray detected but not spectroscopically confirmed, we adopt the photometric redshifts of \citet{Salvato09,Salvato11}, which have been computed for X-ray sources in COSMOS and are optimized for AGN. 

We check to verify that the photometric redshift is in rough agreement with the redshift estimated from visual examination of the source SED. If there is a very large disagreement between the photometric and visual redshift estimates, we adopt the visually estimated redshift. This accounts for occasional failures of the photometric redshift determination due to contaminated photometry in a small number of bands.

\begin{figure}[ht]
       \centering
	\includegraphics[scale=0.5]{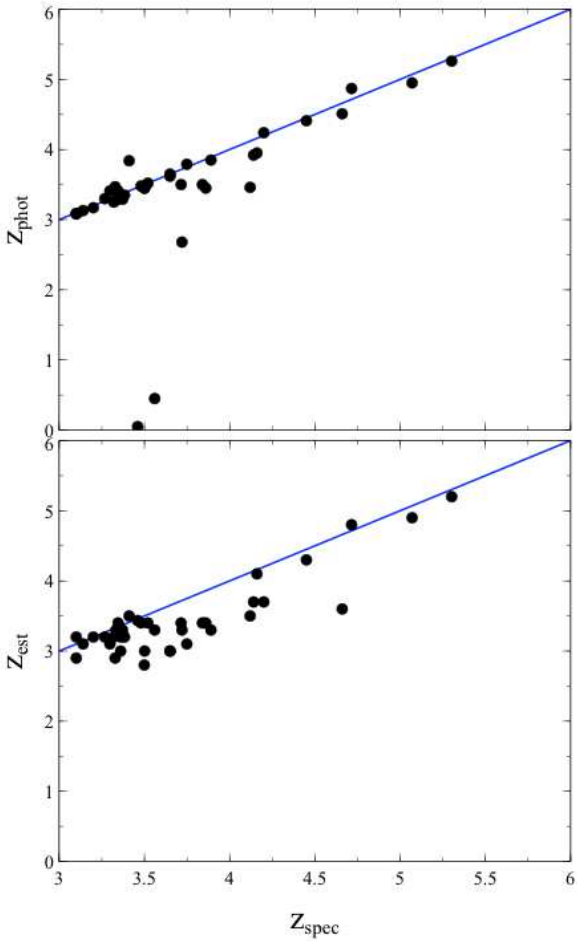}
	\caption{Top: the photometric versus spectroscopic redshift for the sources in our sample with spectroscopic confirmation. The scatter is $\sim$0.7, with a few outliers. Bottom: the estimated redshifts versus spectroscopic redshifts for the same sources. Our visual estimates are systematically low, but with no outliers. These values are only adopted when no spectroscopic redshift is available and the photometric redshift is clearly wrong. This amounts to 29 out of 80 sources in the $3.1 < z < 3.5$ bin and 9 out of 46 sources in the $3.5 < z < 5$ bin.}
\label{Fi:zphotzest}
\end{figure}

In Figure \ref{Fi:zphotzest} we show both the photometric redshift and estimated redshift compared with the spectroscopic redshift for the subset of spectroscopically confirmed candidates. The photometric redshift is generally more accurate, but contains a small number of outliers. The visually estimated redshifts are low on average, but there are no outliers. These estimated redshifts were made by identifying the position of the Lyman-break/Lyman-limit in the source SED. The fact that they are systematically low by a small amount reflects a small bias in this visual estimate. We only adopt these values in cases in which the photometric redshift is likely incorrect, primarily as a result of contaminated photometry (29 of the 80 sources in the $3.1 < z < 3.5$ bin and 9 of the 48 sources in the $3.5 < z < 5$ bin). Because our redshift bins are relatively large, the errors introduced by doing so will not have a significant effect on the derived luminosity functions.

\subsection{Absolute Magnitude Determination}
We follow the convention of giving the quasar luminosity function in terms of the absolute magnitude at rest-frame 1450~\AA. To estimate $M_{1450}$, we fit quasar templates to the source SED, which accounts for elevated fluxes due to quasar emission lines. We found that fitting to the measured magnitudes in the three broad-band Subaru filters $V$, $R$, and $I$ and interpolating to find the observed-frame 1450~\AA\ magnitude often overestimates the luminosity due to the contribution of quasar emission lines to the measured broad-band magnitudes.  

\begin{figure}[ht]
       \centering
	\includegraphics[scale=0.58]{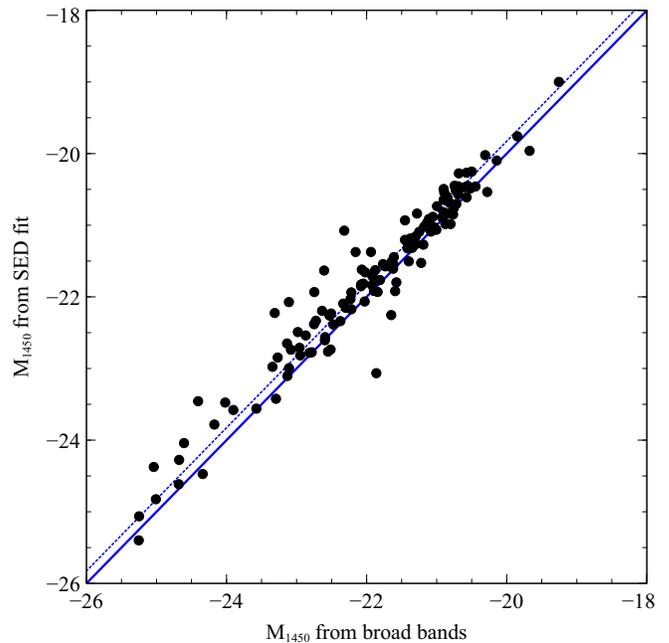}
	\caption{Comparison of the $M_{1450}$ value derived from a linear fit to broad band filters vs. that derived from SED fitting. The absolute magnitude estimated using the SED fit is fainter by $\sim$0.2 magnitudes on average, as indicated with the dotted line. This reflects the fact that the broad band fluxes can be elevated relative to the underlying continuum by broad emission lines.}
\label{Fi:absmagcomparison}
\end{figure}

When performing the SED fits to find $M_{1450}$, only filters within $\pm$2000~\AA\ of the observed-frame wavelength corresponding to rest-frame 1450~\AA\ are considered. The $M_{1450}$ values we find through SED fitting are typically $\sim$0.2 magnitudes fainter than what we found by extrapolating the broad band magnitudes, but with significant scatter (Figure~\ref{Fi:absmagcomparison}).

Given the observed magnitude $m$ corresponding to rest-frame 1450~\AA\ radiation, the absolute magnitude is found with 
\begin{equation} M_{1450} = m - 5 \mathrm{log}(d_{L} / 10) + 2.5 \mathrm{log}(1+z) \end{equation} 
The second term on the right-hand side is the distance modulus; the final term is a correction for the effective narrowing of the frequency interval sampled in the observed frame. 

\subsection{The Luminosity Function at $z\sim3.2$ and $z\sim4$}

\begin{figure}[ht]
        \centering
	\includegraphics[scale=0.58]{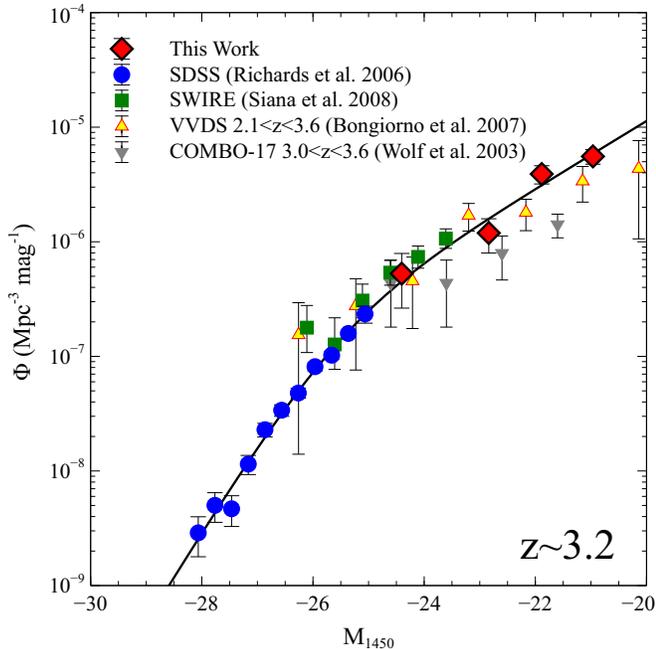}
	\caption{The $z\sim3.2$ luminosity function derived here compared with previous results. The solid black line is our best fit to the double power law parameterization including the bright-end result from the SDSS. Our result is in close agreement with previous determinations of the QLF at similar redshift. The faint-end slope we derive is $\beta = -1.73\pm0.11$.}
\label{Fi:z3.2_lumfunc}
\end{figure}

We compute the luminosity fuction using the derived redshifts and absolute magnitudes for the sample, correcting for the incompleteness at $I\gtrsim24$ from the selection function in Figure~\ref{Fi:completeness3}. The luminosity function we derive for the redshift bin $3.1 < z < 3.5$ is shown in Figure \ref{Fi:z3.2_lumfunc}. There is good continuity with the bright-end result derived in the SDSS, as well as close agreement with prior results that have probed the faint end at similar redshift (e.g. \citealp{Siana08}, \citealp{Bongiorno07}, \citealp{Wolf03}). 

\begin{deluxetable}{cccc}
\tablecolumns{4}
\tablecaption{Binned luminosity function at $z\sim3.2$ and $z\sim4$.}
\tablehead{   
  \colhead{$\Delta M_{1450}$} &
  \colhead{$\mathrm{N}_{\mathrm{QSO}}$} &
  \colhead{$\Phi~(\times10^{-7})$} &
  \colhead{$\Delta\Phi~(\times10^{-7})$}  \\
  \colhead{} &
  \colhead{} &
  \colhead{($\mathrm{Mpc}^{-3}~\mathrm{mag}^{-1}$)} &
  \colhead{} 
}
\startdata
\sidehead{\boldmath$z\sim3.2$\unboldmath} \hline \\
{[}-25.5,-23.5]     & 4 & 5.3 & 2.6 \\  
{[}-23.5,-22.5]     & 9 & 11.9  & 3.9 \\
{[}-22.5,-21.5]     & 26 & 39.0  & 7.1 \\
{[}-21.5,-20.5]     &  41 & 59.8 & 8.4 \\ 
\sidehead{\boldmath$z\sim4$\unboldmath}  \hline \\ 
{[}-25.5,-23.5]     &  5 & 1.9 & 0.8 \\
{[}-23.5,-22.5]     &  9 & 3.5 & 1.2 \\
{[}-22.5,-21.5]     &  14 & 6.4  & 1.4 \\
{[}-21.5,-20.5]     &  20 & 20.6   & 2.6 
\enddata
\end{deluxetable}

We fit the $z\sim$3.2 QLF to the double power law relationship (Equation 2) using our faint-end result in conjunction with the bright end from \citet{Richards06}. The best-fit parameters we derive are $\Phi^{*} = 2.65(\pm2.22)\times10^{-7}$, $M_{1450}^{*} = -25.54\pm0.68$, $\alpha = -2.98\pm0.21$, and $\beta = -1.73\pm0.11$. 

\begin{figure}[ht]
        \centering
	\includegraphics[scale=0.58]{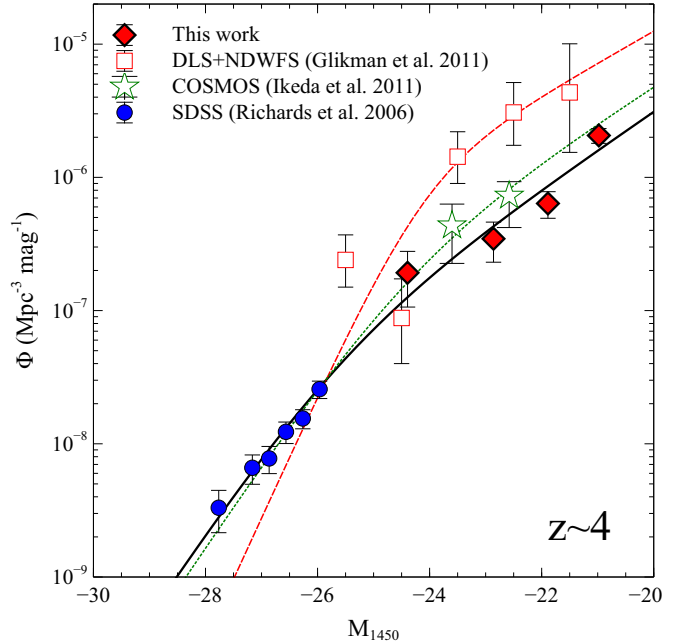}
	\caption{The $z\sim4$ luminosity function derived here compared with previous results. The solid black line is our best fit to the double power law parameterization including the bright-end result from the SDSS. The dotted green line is the best fit from I11 and the dashed red line is the best fit from G11. The disagreement between the bright-end result of G11 and the SDSS data points arises because G11 recompute the SDSS results after finding the absolute magnitudes directly from the SDSS spectra. Our result is in close agreement with that of III. The disagreement in the faint-end space density between the result derived here and that derived in the DLS+NDWFS is significant, with that result higher by a factor of $\sim$4.}

\label{Fi:z4_lumfunc}
\end{figure}

In Figure \ref{Fi:z4_lumfunc} we show the result for the $z\sim4$ luminosity function, together with the previous results from G11 and I11. The best-fit parameters to the double power law paramaterization are $\Phi^{*} = 7.5(\pm23.5)\times10^{-8}$, $M_{1450}^{*} = -25.64\pm2.99$, $\alpha = -2.60\pm0.63$, and $\beta = -1.72\pm0.28$. Our faint end result is in agreement with the result reported by I11, and the faint-end slopes we find at both $z\sim3.2$ and $z\sim4$ agree with the faint-end slope at $2 < z < 3.6$ ($\beta=-1.70$) derived in \citet{Jiang06}. However, the space density of faint quasars at $z\sim4$ we find is lower than the result reported in G11 by a factor of $\sim$3-4. The source of this discrepancy is unclear, but may be due to contamination (dwarf stars and high-redshift galaxies) in the sample of G11 at the faintest magnitudes probed in that work. The point-source selection used in G11 is done with ground-based imaging, which is not as effective as HST imaging at resolving compact, high-z galaxies. Additionally, the number of stellar contaminants in broad band quasar searches grows rapidly at the faintest magnitudes, where spectroscopic follow-up is the most challenging and thus the contamination is not as well-constrained.

\begin{figure}[ht]
        \centering
	\includegraphics[scale=0.58]{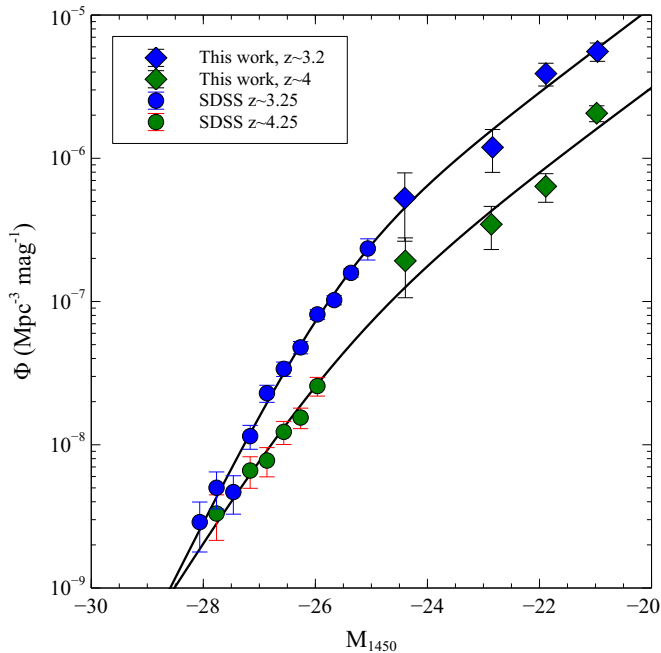}
	\caption{The evolution of the luminosity function from $z$$\sim$4 to $z$$\sim$3.2, combining the bright end derived from the SDSS \citep{Richards06} with the faint end derived here. The space density of faint quasars increases by a factor of $\sim$4 from $z$$\sim$4 to $z$$\sim$3.2. Overplotted are the best-fit lines for each luminosity function using the double power law parameterization. The faint-end slopes we derive for $z$$\sim$3.2 and $z$$\sim$4 are $\beta=-1.73\pm0.11$ and $\beta=-1.72\pm0.28$, respectively.}
\label{Fi:evolution_plot}
\end{figure}

In Figure~\ref{Fi:evolution_plot} we illustrate the strong evolution of the luminosity function from $z\sim3.2$ to $z\sim4$. We find that the space density of faint quasars decreases by a factor of four between $z\sim3.2$ and $z\sim4$, in close analogy with the evolution of brighter optical and X-ray quasars at these redshifts. This trend is also verified by the recent results presented in \citet{Ikeda12}, where it is found that the space density of faint quasars at $z\sim5$ is significantly lower than at $z\sim4$.

\begin{deluxetable}{cccc}
\tablecolumns{4}
\tablecaption{Double power law parameters, combining SDSS and this work.}
\tablehead{   
  \colhead{$\Phi(M^{*})$} &
  \colhead{$M_{1450}^{*}$} &
  \colhead{$\alpha$} &
  \colhead{$\beta$} 
}
\startdata
\sidehead{\boldmath$z\sim3.2\colon$\unboldmath} 
$2.65(\pm2.22)\times10^{-7}$ & $-25.54\pm0.68$ & $-2.98\pm0.21$ & $-1.73\pm0.11$ \\
\sidehead{\boldmath$z\sim4\colon$\unboldmath}   
$7.5(\pm23.5)\times10^{-8}$ & $-25.64\pm2.99$ & $-2.60\pm0.63$ & $-1.72\pm0.28$ 
\enddata
\end{deluxetable}

\section{Possible Sources of Error}

\subsection{Cosmic Variance}


The comoving volume probed here at $z\sim4$ is sufficiently large  ($2.6\times10^{7}~\mathrm{Mpc}^{3}$) that we expect cosmic variance to have little effect. However, it has been shown in the SDSS \citep{Shen07} that the bias of luminous, high redshift quasars relative to dark matter halos increases dramatically with redshift, so we attempt to estimate the likely error introduced by cosmic variance. 

We make use of the cosmic variance cookbook of \citet{Moster11}, in which the bias of galaxies as a function of stellar mass and redshift is used to compute the counting error for galaxies due to cosmic variance. Quasars are expected to be significantly biased with respect to the matter distribution, existing preferentially in the highest mass halos. This leads us to make the approximation that quasars are similarly biased to the most massive ($M \geq 10^{11} M_{\odot}$) galaxies. The bias factor estimated in this way is in relatively good agreement with the results presented in \citet{Shen07} for quasar clustering in the SDSS (and may actually be slightly overestimated).

In the redshift window of $3.5 < z < 5$ in the COSMOS field, we compute an uncertainty of $\sim$19\% in our quasar count, indicating that cosmic variance will contribute only modestly to the derived luminosity function.

\subsection{Contamination by Stars}

The consistent, well-matched wavelength coverage of the COSMOS data, extending from the UV to the infrared, enables the effective  removal of stellar contaminants, the SEDs of which differ substantially from those of quasars. Figure \ref{Fi:ratiotest} illustrates that our final selection preferentially retains objects for which $\chi^{2}$ fitting disfavors stellar templates, and the close agreement with prior studies at $z\sim3$ also indicates that contamination is low. We note that, even it there were significant stellar contamination in the final sample, this would imply that the luminosity functions are upper bounds, making it more difficult to reconcile the result at $z\sim4$ found here with that of G11.

\subsection{Contamination by Star-forming Galaxies}

Because we only have spectroscopic confirmation for a fraction of our sample, we may overestimate the faintest end of the luminosity function by inadvertantly counting compact, star-forming galaxies as quasars. To estimate the  contamination from high-redshift galaxies, we analyze a sample of 386 confirmed star-forming galaxies at  $z > 3$ in COSMOS. These sources were selected as part of a large spectroscopic campaign with Keck DEIMOS (Capak et al., in preparation) aimed at finding high-redshift galaxies down to $I=25$ from $3<z<6$. A number of techniques (Lyman-break criteria, narrow-band selection, etc.) were employed to achieve high completeness, with no selection based on morphology. By applying our quasar selection to these objects, we can get a reasonable sense of the contamination due to compact galaxies. 

Of the 386 confirmed high-redshift galaxies, only 2 (0.5\%) pass our selection (and were removed from our final sample), with the rest rejected primarily on the basis of morphology. We can combine the fraction of high-redshift star forming galaxies likely to contaminate our sample with the UV luminosity function for galaxies at $z\sim4$ \citep{Bouwens07} to estimate the level of contamination. Restricting our attention to the faintest absolute magnitude bin ($M_{1450}=-21$) examined at $z\sim4$, we estimate a contamination level of roughly $10^{-6}$ ($\mathrm{Mpc}^{-3}~\mathrm{mag}^{-1}$) by star-forming galaxies. This is approximately half the space density we derive in this magnitude bin. Therefore, this faintest data point on the luminosity function may be elevated somewhat significantly by star-forming galaxy contamination, leading to an overestimation of the faint-end slope. This issue may need to be settled by follow-up spectroscopy on the faintest sources in the sample.

\subsection{Faint Quasars in Bright Host Galaxies}

Conversely, we may underestimate the faintest bins of the luminosity functions by rejecting bright galaxies hosting faint quasars. This will be the case if a significant fraction of faint type-1 quasars ($M_{1450}\gtrsim-23$) exist in galaxies bright enough in the rest-frame UV to make the source appear extended in ACS $I$-band imaging. Two pieces of evidence lead us to believe that this is not the case.

First, we examine 12 Keck DEIMOS spectra of Chandra sources at high redshift that are classified as extended in the ACS catalog, and find that these sources do not show broad lines indicative of a type-1 quasar. Therefore there is no evidence from spectroscopy of high-redshift X-ray sources of a population of type-1 quasars that we miss by restricting our search to point sources. On the contrary, high-redshift X-ray sources that are morphologically extended seem to be exclusively type-2 AGN. 

Second, we note that, in the sample of $\sim$1400 high-redshift candidate galaxies in COSMOS followed up on with Keck DEIMOS spectroscopy, we do not find cases of broad-line objects with extended morphology. If a significant number of faint type-1 quasars at $z\sim4$ have host galaxies bright enough in the rest-frame UV to make the galaxy appear extended in ACS imaging, we should find serendipitous examples of these in a large sample of candidate high-redshift galaxies. In fact we do not; high-redshift, broad-line emitting objects seem invariably to be point sources in the rest-frame UV. 

We conclude from these two tests that we do not miss a substantial fraction of type-1 quasars due to our morphological criterion, even at the faintest quasar magnitudes. 

\section{Comparison of the X-ray and UV Quasar Luminosity Functions at High Redshift}

The X-ray luminosity function at $z > 3$ in COSMOS was determined recently for the $0.9~\mathrm{deg}^{2}$ Chandra region in COSMOS (\citet{Civano11}, hereafter C11). There it is shown that the population of X-ray luminous AGN decreases rapidly above $z = 3$, in agreement with previous results (e.g. \citealp{Brusa09, Silverman08}). In fact, Figure 4 of C11 shows a decline in space density of X-ray luminous ($L_{\mathrm{x}} \gtrsim 10^{44}~\mathrm{ergs}~\mathrm{s}^{-1}$) AGN by a factor of $\sim$4 between $z\sim3.2$ and $z\sim4$, in agreement with the trend we have demonstrated here for the faint end of the UV QLF at the same redshifts.

\begin{figure}[ht]
        \centering
	\includegraphics[scale=0.32]{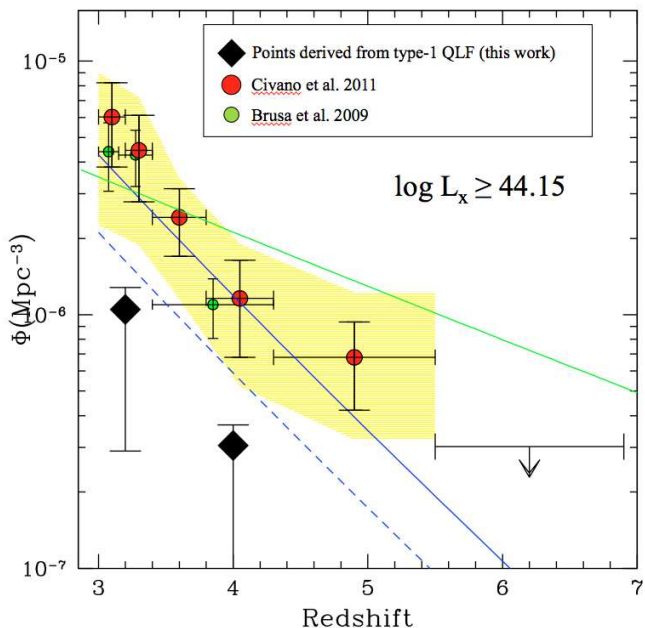}
	\caption{The UV luminosity function results, converted to X-ray (black diamonds), compared to the X-ray luminosity functions for X-ray bright Chandra sources derived in \citet{Civano11} including uncertainties (red circles and yellow shaded area) and for XMM-COSMOS sources (green circles, \citealp{Brusa09}).  The space density is roughly a fourth of that found in X-ray, implying that $\sim$75\% of bright quasars are optically obscured at these redshifts. The blue lines represent the prediction from the \citet{Gilli07} model for obscured and unobscured sources (solid) and unobscured sources only (dashed). }
\label{Fi:xray_bright}
\end{figure}

A rough sense of the obscured fraction at these redshifts can be obtained simply by comparing the high-redshift X-ray sample with our list of type-1 quasars. Doing so, we find that our selection finds 27/101 sources in the high-redshift X-ray sample. The X-ray sources that we miss are primarily classified as extended in the ACS $I$-band imaging, or are fainter than $I=25$, or both. As mentioned previously, an examination of 12 spectra of high-redshift X-ray sources that are classified as extended in ACS imaging shows that they are consistent with being type-2 AGN rather than unobscured quasars. Assuming that the X-ray sources our selection misses are in fact optically obscured, this leads to an estimated type-2 fraction of roughly 75\%. 

To more quantitatitvely compare our results with those found in C11, we convert our rest-frame UV luminosity functions to X-ray using the $\alpha_{ox}$ relationship presented in \citet{Steffen06}$\colon$ 
\begin{equation} 
\resizebox{.9\hsize}{!}{$\mathrm{log}(l_{\mathrm{2 keV}}) = (0.721\pm0.011) \mathrm{log}(l_{\mathrm{2500\AA}}) + (4.531\pm0.688)$} \end{equation} We modify this relationship slightly to put it in terms of the luminosity at 1450~\AA\ by assuming a typical quasar spectral slope.  With this relationship, the space density of quasars with $L_{\mathrm{x}}>10^{44.15}~\mathrm{erg}~\mathrm{s}^{-1}$ is found by integrating the UV luminosity functions over the absolute magnitude interval $M_{1450}=[-\infty, -23.45]$. The uncertainty in the conversion is based on the uncertainty in the parameters of the $\alpha_{ox}$ relation.

Our results are plotted as black diamonds in Figure~\ref{Fi:xray_bright}, together with the X-ray results of C11. The same trend with redshift is observed, but the points derived from the UV QLF are lower by a factor of $\sim$4, in agreement with the ratio found by directly comparing the two samples. While the C11 space density is in agreement with the prediction for obscured and unobscured sources (solid line) derived from the \citet{Gilli07} X-ray background synthesis model, the space density derived from the UV QLF is more consistent with the prediction for unobscured sources only (dashed line). Based on the offset, we again find that the fraction of obscured (but not Compton-thick) quasars is roughly 75\% at $z\sim3-5$.

It is interesting to compare this result with the redshift evolution of the type-2 fraction determined by \citet{Hasinger08} (hereafter H08). Figure~9 of that work clearly indicates an upward trend in obscured fraction with redshift, which seems to saturate around $z=2$ at $\sim$60\%. This result arises after a correction for the fact that observational bias makes it more difficult to identify obscured AGN at higher redshifts. The type-2 fraction we find at $z\sim3-4$, $\sim$75\%, is higher than the result in H08 and may indicate that the trend of increasing obscured fraction with redshift continues to higher redshifts. 

In should also be noted that the result in Figure~9 of H08 is actually a normalization of a rather steep gradient of obscuration versus X-ray luminosity. The obscured fraction decreases with X-ray luminosity at a given redshift, such that for sources above $L_{\mathrm{x}}=10^{44}$ at $z\sim3-4$ the estimated obscured fraction is actually closer to 20-30\%, significantly lower than what we find for comparably luminous sources at these redshifts. Follow-up spectroscopy on more of the $z>3$ X-ray sources in COSMOS would allow a more firm determination of the type-2 fraction at these redshifts.

\section{The Ionizing Background Due to Quasars}

Given the luminosity function $\Phi(M_{1450},z)$, we can estimate the rate of production of ionizing photons by quasars at redshift $z$. To do so, we use results from \citet{Vanden01} and \citet{Telfer02} to approximate the spectral energy distribution of a typical quasar: \begin{equation}
L(\nu)\propto \left\{ \begin{array}{ll}
 \nu^{-0.5} & \mbox{($1216~\mathrm{\AA} < \lambda < 2500~\mathrm{\AA})$} \\
\nu^{-1.76} & \mbox{$(\lambda < 1216~\mathrm{\AA})$} \\
\end{array}
\right. \\
\end{equation}

The number of hydrogen-ionizing photons emitted per second by a quasar of absolute magnitude $M_{1450}$ is given by \begin{equation} \dot{n}_{\mathrm{HI}}(M_{1450}) = \int_{\mathrm{1 ryd}}^{\mathrm{4 ryd}} \frac{L(M_{1450},\nu)}{h\nu}\,d\nu \end{equation} where the cutoff at 4 ryd is due to preferential HeII absorption of photons above this energy \citep{Madau99}. Similarly, the rate of emission of HeII ionizing photons is given by \begin{equation} \dot{n}_{\mathrm{HeII}}(M_{1450}) = \int_{\mathrm{4 ryd}}^{\mathrm{\infty}} \frac{L(M_{1450},\nu)}{h\nu}\,d\nu \end{equation}

Combining equations (5) and (6) with the QLF, one can compute the number of HI and HeII ionizing photons produced per second per unit comoving volume at a given redshift: \begin{equation} \dot{\mathcal{N}}_{\mathrm{HI}}(z) = \int \Phi(M,z) \dot{n}_{\mathrm{HI}}(M)\,dM \end{equation} \begin{equation} \dot{\mathcal{N}}_{\mathrm{HeII}}(z) = \int \Phi(M,z) \dot{n}_{\mathrm{HeII}}(M)\,dM \end{equation}

We use the results for the QLF at $z\sim3.2$ and $z\sim4$ presented in \S~7 to estimate these ionizing backgrounds and their evolution with redshift.

\subsection{HI Ionizing Background}
With the QLFs derived in \S~7, we find quasar emission rates of hydrogen-ionizing photons at $z\sim3.2$ and $z\sim4$ of $\dot{\mathcal{N}}_{\mathrm{HI}}=3.3\times10^{50}~\mathrm{s}^{-1}~\mathrm{Mpc}^{-3}$ and $1.0\times10^{50}~\mathrm{s}^{-1}~\mathrm{Mpc}^{-3}$, respectively. These rates are lower than needed to reionize the IGM at either redshift. At $z$=4, for example, the estimated required rate for reionization is $\dot{\mathcal{N}}_{\mathrm{HI}} =2.4\times10^{51}~\mathrm{s}^{-1}~\mathrm{Mpc}^{-3}$ \citep{Madau99}, more than 20 times higher than what we find. In addition, the rates we compute ignore escape fraction and are therefore upper limits on the photons available to ionize the IGM. 

The emissivity of UV radiation due to quasars can be found with our luminosity functions and compared with the result given in \citet{Haardt12}. We integrate to find the emissivity at 1450~\AA\ and use a conversion factor of $\mathrm{f}_{912\mathrm{\AA}}$ / $\mathrm{f}_{1450\mathrm{\AA}}$ = 0.58 to convert to the corresponding emissivity at 912~\AA. At $z\sim3.2$ and $z\sim4$ we find $\epsilon_{\mathrm{912\AA}} = 4.4\times10^{24}$ and $1.4\times10^{24}~\mathrm{ergs}~\mathrm{s}^{-1}~\mathrm{Mpc}^{-3}~\mathrm{Hz}^{-1}$, respectively. \citet{Haardt12} find $\epsilon_{\mathrm{912\AA}} = 4.3\times10^{24}$ and $2.0\times10^{24}~\mathrm{ergs}~\mathrm{s}^{-1}~\mathrm{Mpc}^{-3}~\mathrm{Hz}^{-1}$ at the same redshifts. We are in relatively close agreement, although our value at $z\sim4$ is somewhat lower, likely due to the rapid decline in the space density of faint quasars we find. 

 The emissivity we find at $z\sim4$ is significantly lower than that required to maintain reionization at $z\sim7$ \citep{Robertson10}, even assuming a very high escape fraction for ionizing photons. Therefore, assuming that the faint-end of the QLF continues to decline at higher redshifts, type-1 quasars can not contribute significantly to cosmic reionization of hydrogen at $z\sim6-10$. 

\subsection{HeII Ionizing Background}
We find a QSO emission rate of HeII-ionizing photons at $z\sim3.2$ of $\dot{\mathcal{N}}_{\mathrm{HeII}}=3.1\times10^{49}~\mathrm{s}^{-1}~\mathrm{Mpc}^{-3}$. At $z\sim4$ we find $\dot{\mathcal{N}}_{\mathrm{HeII}}=1.0\times10^{49}~\mathrm{s}^{-1}~\mathrm{Mpc}^{-3}$. The significant increase in the production by quasars of HeII-ionizing ($>$54.4 eV) photons from $z\sim4\rightarrow3$ is qualitatively consistent with the observed patchy reionization of HeII at $z\sim3$  \citep{Shull10, Bolton06}. While AGN are considered the only sources capable of producing sufficient photons with energies above 54.4 eV to be responsible for the reionization of HeII, significant uncertainty exists with regard to the specifics of the HeII reionization process \citep{Shull10}. This is mainly due to degeneracies between temperature, density, and ionization fraction in the IGM, as well as the dependence on the spatial distribution and spectral hardness of the ionizing sources. Therefore, we do not attempt to quantify the ionizing background needed to produce the observed HeII reionization at $z\sim3$.

\section{Summary \& Conclusions}

We have searched the COSMOS field for type-1 quasars at $3.1 < z < 5$ and $I_{\mathrm{AB}} < 25$, with care taken to achieve high completeness. Our method has been to apply a weak quasar selection that is nearly complete but unreliable,  and then identify and remove contaminants through careful inspection of the COSMOS photometric and imaging data. This method exploits the fact that stellar contaminants are easily identified with the 29 bands of deep photometry spanning the UV to the infrared. We recover all 39 previously known type-1 quasars at $z>3.1$ in COSMOS as well as 116 additional likely quasars, and have presented evidence based on simulations and automated classifiers that this sample is highly complete for type-1 quasars above $z=3.1$ in the HST-ACS region (1.64 $\mathrm{deg}^{2}$) of COSMOS.  

Here we summarize our main conclusions.

\begin{enumerate}
\item{We find 155 likely type-1 quasars at $z > 3.1$ in the COSMOS field, 39 of which have been previously confirmed, down to a limiting magnitude $I_{AB} = 25$. We present strong evidence that this quasar sample is nearly complete over the redshift range $3.1 < z < 5$.}
\item{We use the quasar sample to compute the QLF at $z\sim3.2$ and $z\sim4$. The faint-end results we obtain show continuity with the bright-end results reported by \citet{Richards06} for the SDSS, and demonstrate a clear decrease in the space density of faint quasars, by roughly a factor of four, from $z\sim3.2$ to $z\sim4$. The faint-end we derive at $z\sim4$ is in good agreement with the result reported for the COSMOS field in I11, but disagrees with the result reported in G11. While the source of this discrepancy is unclear, we suggest that it may be caused by contaminating stars/high-redshift galaxies at the faintest magnitudes in the sample of G11.} 
\item{We find no evolution in the faint-end slope $\beta$ over the redshift range investigated. The faint end slopes we find are $\beta=-1.73\pm0.11$ at $z\sim3.2$ and $\beta=-1.72\pm0.28$ at $z\sim4$, similar to those found at lower redshift.}
\item{We compare the luminosity functions derived here with the X-ray luminosity function at $z>3$ in COSMOS. The optical QLF and the X-ray QLF evolve similarly between $z\sim3.2$ and $z\sim4$ in the sense that both show a rapid decline in space density. However, the different normalizations of the LFs imply that roughly 75\% of AGN with $\mathrm{log}(L_{x}) \gtrsim 43.75$ are obscured, type-2 quasars at $z\sim3-4$.}

\item{We compute the ionizing background due to quasars, both of HeII and HI. The rapid increase in production of HeII ionizing photons from $z\sim4\rightarrow3$ that we find is qualitatively consistent with the observed onset of HeII reionization at $z\sim3$. However, given the decline of the faint end of the QLF to higher redshift, we conclude that faint quasars are unlikely to contribute substantially to cosmic reionization of hydrogen.}
\end{enumerate}

\acknowledgments
We gratefully acknowledge the contributions of the entire COSMOS collaboration. More information about the COSMOS survey is available at http://www.astro.caltech.edu/∼cosmos. We acknowledge the anonymous referee for suggestions that significantly improved this work. We thank Dr. Eilat Glikman for helpful discussions.  M.S. and G.H. acknowledge support by the German Deutsche Forschungsgemeinschaft, DFG Leibniz Prize (FKZ HA 1850/28-1). D.M. gratefully acknowledges the support of the Infrared Processing and Analysis Center (IPAC) Visiting Graduate Fellowship program, as well as the Carnegie Visiting Graduate Fellowship program.

\bibliographystyle{apj}
\bibliography{qsopapers}

\appendix
\begin{center}
\setlength{\tabcolsep}{11pt}

\begin{longtable*}{cccccccccccc}
\caption[Summary of the confirmed and likely quasars found here and used to compute the luminosity fuction.]{Summary of the likely quasars found here and used to compute the luminosity function. We give $z_{est}$, the redshift we estimated based on visual examination, $z_{qso}$, the photometric redshift from COSMOS, $z_{spec}$, the spectroscopic redshift, if it exists, and $z_{used}$, the redshift we adopt in computing the luminosity function. In addition, we list the confidence flag we assigned the source based on visual examination, whether or not the source is detected in X-ray, and the absolute magnitude $M_{1450}$ we derive.} \\
\label{Ta:quasars}  \\

\hline \hline \\[-2ex]
   \multicolumn{1}{c}{\textbf{ID}} &
   \multicolumn{1}{c}{\textbf{RA}} &
   \multicolumn{1}{c}{\textbf{DEC}} &
   \multicolumn{1}{c}{\textbf{$I_{AB}$}} &
   \multicolumn{1}{c}{\textbf{$z_{est}$}} &
   \multicolumn{1}{c}{\textbf{$z_{qso}$}} &  
   \multicolumn{1}{c}{\textbf{$z_{spec}$}} &
   \multicolumn{1}{c}{\textbf{$z_{used}$}} &
   \multicolumn{1}{c}{\textbf{Flag}} &
   \multicolumn{1}{c}{\textbf{X-ray}} &
  \multicolumn{1}{c}{\textbf{$M_{1450}$}} \\ 

   \textbf{} & \textbf{(J2000)} & \textbf{(J2000)} & \textbf{(AUTO)} & \textbf{} & \textbf{} & \textbf{} & \textbf{} & \textbf{} & \textbf{} & \textbf{} & \textbf{} \\ [0.5ex] \hline

   \\[-1.8ex] 

\endfirsthead

\multicolumn{3}{c}{{\tablename} \thetable{} -- Continued} \\[0.5ex]
  \hline \hline \\[-2ex]
   \multicolumn{1}{c}{\textbf{ID}} &
   \multicolumn{1}{c}{\textbf{RA}} &
   \multicolumn{1}{c}{\textbf{DEC}} &
   \multicolumn{1}{c}{\textbf{$I_{AB}$}} &
   \multicolumn{1}{c}{\textbf{$z_{est}$}} &
   \multicolumn{1}{c}{\textbf{$z_{qso}$}} &  
   \multicolumn{1}{c}{\textbf{$z_{spec}$}} &
   \multicolumn{1}{c}{\textbf{$z_{used}$}} &
   \multicolumn{1}{c}{\textbf{Flag}} &
   \multicolumn{1}{c}{\textbf{X-ray}} &
  \multicolumn{1}{c}{\textbf{$M_{1450}$}} \\ 

   \textbf{} & \textbf{(J2000)} & \textbf{(J2000)} & \textbf{(AUTO)} & \textbf{} & \textbf{} & \textbf{} & \textbf{} & \textbf{} & \textbf{} & \textbf{} & \textbf{} \\ [0.5ex] 
 
  \\[-1.8ex]
\endhead

  \multicolumn{3}{l}{{Continued on Next Page\ldots}} \\
\endfoot

  \\[-1.8ex] \hline
\endlastfoot

381470 &       149.85396 &        1.753672 & 22.55 &  3.30 &  0.11 &  -99 &   3.30 &       4 & no & -22.11 \\ \hline
1625825 &       150.43993 &        2.703496 & 23.05 &  3.43 &  0.05 &  3.46 &   3.46 &       4 & yes & -21.99 \\ \hline
1199385 &       150.39026 &        2.445338 & 24.07 &  3.18 &  2.78 &  -99 &   3.18 &       4 & no & -20.60 \\ \hline
1575750 &       150.73715 &        2.722578 & 22.23 &  3.20 &  3.25 &  3.32 &   3.32 &       4 & yes & -22.78 \\ \hline
910592 &       150.73557 &        2.199578 & 20.22 &  3.00 &  3.45 &  3.50 &   3.50 &       4 & yes & -25.06 \\ \hline
1163086 &       150.70377 &        2.370019 & 22.44 &  3.10 &  3.79 &  3.75 &   3.75 &       4 & yes & -23.00 \\ \hline
1159815 &       150.63844 &        2.391350 & 22.16 &  3.00 &  3.62 &  3.65 &   3.65 &       4 & yes & -22.98 \\ \hline
1605275 &       150.62006 &        2.671402 & 21.83 &  3.10 &  3.13 &  3.14 &   3.14 &       4 & yes & -22.82 \\ \hline
887716 &       149.49590 &        1.968019 & 22.61 &  3.30 &  3.23 &  -99 &   3.23 &       4 & no & -22.38 \\ \hline
1381605 &       150.57439 &        2.552482 & 23.60 &  3.20 &  3.13 &  -99 &   3.13 &       4 & no & -21.54 \\ \hline
503666 &       150.55476 &        1.904693 & 23.90 &  3.00 &  3.33 &  -99 &   3.33 &       4 & no & -21.22 \\ \hline
299482 &       150.49628 &        1.638000 & 23.99 &  3.30 &  3.35 &  -99 &   3.35 &       4 & no & -21.31 \\ \hline
1856470 &       150.47568 &        2.798362 & 21.28 &  3.50 &  3.46 &  4.12 &   4.12 &       4 & no & -24.37 \\ \hline
507779 &       150.48563 &        1.871927 & 22.03 &  4.30 &  4.41 &  4.45 &   4.45 &       4 & no & -23.78 \\ \hline
739700 &       150.45497 &        1.967424 & 24.03 &  3.40 &  3.48 &  3.48 &   3.48 &       4 & yes & -21.20 \\ \hline
504209 &       150.42476 &        1.900266 & 24.79 &  3.40 &  1.18 &  -99 &   3.40 &       4 & yes & -20.10 \\ \hline
748601 &       150.38388 &        2.074549 & 23.46 &  3.50 &  3.45 &  -99 &   3.45 &       4 & yes & -21.76 \\ \hline
1208399 &       150.25954 &        2.376141 & 21.42 &  3.30 &  2.68 &  3.72 &   3.72 &       4 & yes & -24.47 \\ \hline
1657280 &       150.24078 &        2.659058 & 22.33 &  3.00 &  3.33 &  3.36 &   3.36 &       4 & yes & -22.74 \\ \hline
1448618 &       150.21497 &        2.582674 & 24.52 &  5.20 &  5.26 &  5.30 &   5.30 &       4 & yes & -22.73 \\ \hline
1463661 &       150.20885 &        2.481935 & 20.11 &  3.30 &  3.28 &  3.33 &   3.33 &       4 & yes & -25.40 \\ \hline
1224733 &       150.20898 &        2.438466 & 21.15 &  3.40 &  3.50 &  3.71 &   3.71 &       4 & yes & -24.28 \\ \hline
560579 &       150.21025 &        1.853839 & 22.12 &  3.40 &  3.45 &  -99 &   3.45 &       4 & yes & -23.10 \\ \hline
1231613 &       150.21075 &        2.391473 & 22.55 &  2.90 &  3.08 &  3.10 &   3.10 &       4 & yes & -22.33 \\ \hline
565710 &       150.21199 &        1.818717 & 24.08 &  3.50 &  3.48 &  -99 &   3.48 &       4 & no & -21.18 \\ \hline
348143 &       150.17534 &        1.649839 & 24.19 &  3.30 &  3.28 &  -99 &   3.28 &       4 & no & -20.65 \\ \hline
1900976 &       150.15880 &        2.808512 & 23.06 &  3.50 &  0.83 &  -99 &   3.50 &       4 & no & -21.58 \\ \hline
1465836 &       150.13036 &        2.466012 & 23.04 &  3.40 &  3.45 &  3.86 &   3.86 &       4 & no & -22.56 \\ \hline
1221992 &       150.13364 &        2.457429 & 24.14 &  3.10 &  3.13 &  -99 &   3.13 &       4 & yes & -20.74 \\ \hline
1226535 &       150.10098 &        2.419435 & 22.33 &  3.60 &  4.51 &  4.66 &   4.66 &       4 & yes & -23.46 \\ \hline
330806 &       150.10738 &        1.759201 & 22.55 &  3.70 &  3.92 &  4.14 &   4.14 &       4 & yes & -22.85 \\ \hline
780598 &       150.09685 &        2.021495 & 24.33 &  3.50 &  3.23 &  -99 &   3.23 &       4 & yes & -20.49 \\ \hline
582598 &       150.04268 &        1.872161 & 23.82 &  3.30 &  3.33 &  3.36 &   3.36 &       4 & yes & -21.80 \\ \hline
585189 &       150.00940 &        1.852637 & 23.67 &  3.00 &  3.33 &  -99 &   3.33 &       4 & yes & -21.92 \\ \hline
804307 &       150.00438 &        2.038898 & 21.86 &  2.80 &  3.45 &  3.50 &   3.50 &       4 & yes & -23.56 \\ \hline
1261211 &       149.90547 &        2.354014 & 23.19 &  3.20 &  3.30 &  3.27 &   3.27 &       4 & no & -21.92 \\ \hline
1054048 &       149.87920 &        2.225839 & 22.70 &  3.00 &  3.65 &  3.65 &   3.65 &       4 & yes & -22.77 \\ \hline
1249763 &       149.89417 &        2.432972 & 23.29 &  3.20 &  3.35 &  3.38 &   3.38 &       4 & yes & -21.84 \\ \hline
1271385 &       149.86966 &        2.294046 & 21.67 &  3.40 &  3.41 &  3.35 &   3.35 &       4 & yes & -23.42 \\ \hline
1046585 &       149.85153 &        2.276400 & 23.02 &  3.30 &  3.29 &  3.37 &   3.37 &       4 & yes & -22.39 \\ \hline
1511846 &       149.84576 &        2.481679 & 23.11 &  3.20 &  3.33 &  3.36 &   3.36 &       4 & yes & -21.94 \\ \hline
1513806 &       149.78207 &        2.471342 & 23.70 &  3.40 &  3.26 &  -99 &   3.26 &       4 & yes & -20.88 \\ \hline
1719143 &       149.75539 &        2.738555 & 22.87 &  3.40 &  3.52 &  3.52 &   3.52 &       4 & yes & -22.26 \\ \hline
1272246 &       149.78381 &        2.452135 & 23.70 &  4.90 &  4.95 &  5.07 &   5.07 &       4 & yes & -22.23 \\ \hline
1273346 &       149.77692 &        2.444306 & 22.78 &  4.10 &  3.95 &  4.16 &   4.16 &       4 & yes & -22.65 \\ \hline
1284334 &       149.77104 &        2.365819 & 24.64 &  3.60 &  3.45 &  -99 &   3.45 &       4 & yes & -20.25 \\ \hline
1060679 &       149.73622 &        2.179933 & 23.45 &  3.70 &  4.24 &  4.20 &   4.20 &       4 & yes & -22.23 \\ \hline
1720201 &       149.74873 &        2.732016 & 23.95 &  3.10 &  3.15 &  -99 &   3.15 &       4 & yes & -20.54 \\ \hline
422327 &       149.70151 &        1.638375 & 22.41 &  3.20 &  3.17 &  3.20 &   3.20 &       4 & no & -22.49 \\ \hline
1743444 &       149.66605 &        2.740230 & 22.51 &  3.20 &  3.15 &  -99 &   3.15 &       4 & no & -22.54 \\ \hline
1551171 &       149.52461 &        2.531040 & 22.74 &  3.30 &  3.27 &  -99 &   3.27 &       4 & no & -21.71 \\ \hline
1330271 &       149.52908 &        2.380164 & 20.80 &  3.20 &  3.09 &  3.10 &   3.10 &       4 & yes & -24.04 \\ \hline
699705 &       150.58488 &        2.081361 & 23.12 &  3.00 &  3.29 &  -99 &   3.29 &       4 & yes & -21.83 \\ \hline
1371806 &       150.59184 &        2.619375 & 22.83 &  3.00 &  3.12 &  -99 &   3.12 &       3 & no & -22.10 \\ \hline
1628943 &       150.55078 &        2.682909 & 23.91 &  3.30 &  0.45 &  3.56 &   3.56 &       3 & no & -21.31 \\ \hline
361333 &       149.92613 &        1.724807 & 23.46 &  3.10 &  3.41 &  3.30 &   3.30 &       3 & no & -21.57 \\ \hline
1518518 &       149.69594 &        2.603082 & 24.24 &  3.30 &  3.29 &  -99 &   3.29 &       3 & no & -20.64 \\ \hline
388046 &       149.72974 &        1.704067 & 23.62 &  3.30 &  0.17 &  -99 &   3.30 &       3 & no & -21.32 \\ \hline
1420590 &       150.36694 &        2.616221 & 23.65 &  3.30 &  3.83 &  -99 &   3.30 &       3 & no & -21.37 \\ \hline
990122 &       150.29726 &        2.148846 & 20.46 &  2.90 &  3.47 &  3.33 &   3.33 &       3 & yes & -24.62 \\ \hline
1648871 &       150.31337 &        2.716225 & 24.70 &  3.70 &  3.63 &  -99 &   3.63 &       3 & no & -20.52 \\ \hline
530538 &       150.27693 &        1.885094 & 24.08 &  4.80 &  4.87 &  4.72 &   4.72 &       3 & no & -22.20 \\ \hline
599355 &       149.85320 &        1.920494 & 22.66 &  3.40 &  0.74 &  -99 &   3.40 &       3 & no & -22.17 \\ \hline
1870817 &       150.24617 &        2.858176 & 23.82 &  3.40 &  3.55 &  -99 &   3.55 &       3 & no & -21.45 \\ \hline
1454738 &       150.20822 &        2.539911 & 22.46 &  3.30 &  3.24 &  -99 &   3.24 &       3 & no & -21.55 \\ \hline
618000 &       149.83923 &        1.793837 & 24.63 &  3.40 &  3.55 &  -99 &   3.55 &       3 & no & -20.77 \\ \hline
1730531 &       149.84322 &        2.659095 & 22.90 &  3.20 &  3.51 &  -99 &   3.51 &       3 & no & -22.15 \\ \hline
1970813 &       149.64572 &        2.797057 & 23.24 &  3.40 &  4.15 &  -99 &   3.40 &       3 & no & -21.63 \\ \hline
115356 &       150.14893 &        1.587745 & 24.06 &  3.80 &  3.88 &  -99 &   3.88 &       3 & no & -21.50 \\ \hline
1717612 &       149.76262 &        2.749090 & 24.81 &  3.90 &  0.49 &  -99 &   3.90 &       3 & no & -20.98 \\ \hline
329051 &       150.16891 &        1.774590 & 23.24 &  4.30 &  4.35 &  -99 &   4.35 &       3 & no & -22.71 \\ \hline
790476 &       150.11644 &        1.963943 & 23.90 &  3.50 &  3.84 &  3.41 &   3.41 &       3 & no & -21.25 \\ \hline
1326857 &       149.55511 &        2.402630 & 24.32 &  3.30 &  3.94 &  -99 &   3.94 &       3 & no & -21.53 \\ \hline
803073 &       150.05769 &        2.046285 & 24.88 &  3.16 &  0.00 &  -99 &   3.16 &       3 & no & -21.51 \\ \hline
373838 &       150.02069 &        1.639405 & 24.74 &  3.34 &  3.20 &  -99 &   3.20 &       3 & no & -20.43 \\ \hline
374612 &       149.91289 &        1.632472 & 24.07 &  3.12 &  0.06 &  -99 &   3.12 &       3 & no & -19.52 \\ \hline
1727644 &       149.85581 &        2.681164 & 24.47 &  3.16 &  0.12 &  -99 &   3.16 &       3 & no & -20.71 \\ \hline
376803 &       149.85549 &        1.768740 & 21.06 &  3.30 &  0.09 &  -99 &   3.30 &       3 & no & -20.20 \\ \hline
1283260 &       149.78825 &        2.372171 & 23.38 &  3.40 &  0.24 &  -99 &   3.40 &       3 & no & -21.61 \\ \hline
594007 &       149.85205 &        1.959338 & 24.39 &  3.26 &  3.35 &  -99 &   3.35 &       3 & no & -20.59 \\ \hline
1155791 &       150.66118 &        2.414451 & 24.05 &  3.22 &  3.15 &  -99 &   3.15 &       3 & no & -20.97 \\ \hline
1153009 &       150.58006 &        2.439344 & 24.23 &  3.16 &  3.10 &  -99 &   3.10 &       3 & no & -20.78 \\ \hline
842269 &       149.58212 &        2.117000 & 23.36 &  3.16 &  0.11 &  -99 &   3.16 &       3 & no & -21.62 \\ \hline
851569 &       149.56419 &        2.055025 & 24.68 &  3.50 &  3.75 &  -99 &   3.75 &       3 & no & -20.84 \\ \hline
973691 &       150.34480 &        2.257383 & 24.27 &  5.08 &  4.95 &  -99 &   4.95 &       3 & no & -21.55 \\ \hline
1200307 &       150.28623 &        2.424521 & 23.34 &  3.10 &  1.18 &  -99 &   3.10 &       3 & no & -20.96 \\ \hline
1675730 &       150.07668 &        2.702570 & 23.94 &  3.10 &  0.05 &  -99 &   3.10 &       3 & no & -21.07 \\ \hline
1234913 &       150.11537 &        2.363513 & 23.06 &  3.20 &  3.46 &  -99 &   3.46 &       3 & no & -21.03 \\ \hline
1413401 &       150.47971 &        2.496383 & 24.24 &  3.12 &  0.00 &  -99 &   3.12 &       3 & no & -20.65 \\ \hline
1897688 &       150.07535 &        2.828791 & 23.78 &  3.50 &  0.61 &  -99 &   3.50 &       3 & no & -21.27 \\ \hline
502784 &       150.40532 &        1.910871 & 24.09 &  3.12 &  2.62 &  -99 &   3.12 &       3 & no & -20.44 \\ \hline
754632 &       150.38487 &        2.036528 & 24.15 &  3.12 &  0.41 &  -99 &   3.12 &       3 & no & -20.79 \\ \hline
1464601 &       150.21416 &        2.475017 & 23.19 &  3.11 &  0.81 &  -99 &   3.11 &       3 & yes & -21.71 \\ \hline
1498712 &       149.76903 &        2.573805 & 23.41 &  3.14 &  0.24 &  -99 &   3.14 &       3 & no & -21.28 \\ \hline
1110682 &       149.50595 &        2.185332 & 22.01 &  3.28 &  0.09 &  -99 &   3.28 &       3 & no & -22.71 \\ \hline
1781388 &       149.50613 &        2.642649 & 24.94 &  3.72 &  1.40 &  -99 &   3.72 &       3 & no & -20.30 \\ \hline
265707 &       150.70856 &        1.698028 & 23.86 &  3.40 &  0.17 &  -99 &   3.40 &       3 & no & -21.03 \\ \hline
593446 &       150.05812 &        1.794438 & 24.44 &  3.20 &  3.16 &  -99 &   3.16 &       3 & no & -20.45 \\ \hline
644212 &       149.53085 &        1.953905 & 24.28 &  3.40 &  0.21 &  -99 &   3.40 &       3 & no & -20.49 \\ \hline
1043168 &       150.00752 &        2.132265 & 24.25 &  3.40 &  3.33 &  -99 &   3.33 &       3 & no & -20.83 \\ \hline
259947 &       150.61942 &        1.737010 & 24.60 &  3.10 &  3.13 &  -99 &   3.13 &       3 & no & -20.46 \\ \hline
878986 &       149.47784 &        2.036321 & 24.76 &  4.10 &  4.00 &  -99 &   4.00 &       3 & no & -20.61 \\ \hline
262808 &       150.63403 &        1.719672 & 24.96 &  4.60 &  4.46 &  -99 &   4.46 &       3 & no & -20.84 \\ \hline
710344 &       150.62828 &        2.006204 & 23.36 &  3.60 &  3.45 &  -99 &   3.45 &       3 & no & -22.06 \\ \hline
1384622 &       150.59956 &        2.534162 & 23.29 &  3.20 &  0.53 &  -99 &   3.20 &       3 & no & -21.28 \\ \hline
702684 &       150.56259 &        2.060279 & 24.34 &  3.70 &  3.70 &  -99 &   3.70 &       3 & no & -20.98 \\ \hline
1255908 &       149.90977 &        2.391198 & 24.69 &  3.90 &  3.85 &  -99 &   3.85 &       3 & no & -20.85 \\ \hline
1723659 &       149.77576 &        2.704966 & 17.94 &  3.70 &  3.77 &  -99 &   3.77 &       2 & no & -22.76 \\ \hline
428704 &       149.55103 &        1.764459 & 24.59 &  3.10 &  3.20 &  -99 &   3.20 &       2 & no & -20.47 \\ \hline
462793 &       150.73788 &        1.884472 & 24.97 &  3.20 &  1.05 &  -99 &   3.20 &       2 & no & -19.00 \\ \hline
1041149 &       150.00462 &        2.143163 & 23.88 &  3.22 &  1.04 &  -99 &   3.22 &       2 & no & -22.82 \\ \hline
1257886 &       149.97707 &        2.378910 & 24.71 &  3.12 &  0.21 &  -99 &   3.12 &       2 & no & -20.50 \\ \hline
1558001 &       149.50140 &        2.469177 & 24.66 &  5.12 &  5.42 &  -99 &   5.42 &       2 & no & -22.67 \\ \hline
1309325 &       149.56500 &        2.362180 & 24.65 &  5.60 &  5.28 &  -99 &   5.28 &       2 & no & -23.07 \\ \hline
1260818 &       149.92712 &        2.358468 & 24.16 &  3.20 &  3.44 &  -99 &   3.44 &       2 & no & -21.06 \\ \hline
405715 &       149.70297 &        1.752171 & 24.95 &  3.10 &  3.38 &  -99 &   3.38 &       2 & no & -20.27 \\ \hline
766066 &       150.06465 &        2.124108 & 24.51 &  5.00 &  4.85 &  -99 &   4.85 &       2 & no & -21.93 \\ \hline
1631571 &       150.51112 &        2.665648 & 24.61 &  3.00 &  3.82 &  -99 &   3.82 &       2 & no & -21.09 \\ \hline
1383442 &       150.67668 &        2.543255 & 23.28 &  3.20 &  0.48 &  -99 &   3.20 &       2 & no & -21.09 \\ \hline
361093 &       149.92401 &        1.727810 & 23.83 &  3.33 &  0.67 &  -99 &   3.33 &       2 & no & -20.86 \\ \hline
394142 &       149.82733 &        1.665374 & 23.40 &  3.12 &  0.25 &  -99 &   3.12 &       2 & no & -21.53 \\ \hline
510917 &       150.49745 &        1.851680 & 24.16 &  4.60 &  4.39 &  -99 &   4.39 &       2 & no & -21.82 \\ \hline
380027 &       149.88055 &        1.761200 & 24.01 &  3.60 &  0.63 &  -99 &   3.60 &       2 & no & -21.02 \\ \hline
298002 &       150.43706 &        1.649305 & 23.35 &  3.30 &  3.85 &  3.89 &   3.89 &       2 & no & -22.34 \\ \hline
203339 &       149.51596 &        1.609118 & 23.76 &  3.20 &  3.49 &  -99 &   3.49 &       2 & no & -20.85 \\ \hline
964908 &       150.42227 &        2.150364 & 23.97 &  3.30 &  0.28 &  -99 &   3.30 &       2 & no & -20.89 \\ \hline
1338373 &       149.48837 &        2.327380 & 24.52 &  5.10 &  4.64 &  -99 &   4.64 &       2 & no & -21.66 \\ \hline
598191 &       149.80116 &        1.927920 & 24.20 &  4.60 &  4.36 &  -99 &   4.36 &       2 & no & -21.75 \\ \hline
232585 &       150.73338 &        1.781481 & 24.11 &  3.20 &  0.22 &  -99 &   3.20 &       2 & no & -20.91 \\ \hline
535642 &       150.30261 &        1.852066 & 24.46 &  3.40 &  3.50 &  3.84 &   3.84 &       2 & no & -20.91 \\ \hline
1385272 &       150.66348 &        2.529267 & 23.65 &  3.30 &  3.12 &  -99 &   3.12 &       2 & no & -21.16 \\ \hline
519634 &       150.27715 &        1.958373 & 22.14 &  3.40 &  0.70 &  -99 &   3.40 &       2 & no & -22.61 \\ \hline
1110702 &       149.55306 &        2.185063 & 23.20 &  3.44 &  3.46 &  -99 &   3.46 &       2 & no & -20.94 \\ \hline
1432719 &       150.26303 &        2.520890 & 24.22 &  3.30 &  0.52 &  -99 &   3.30 &       2 & no & -20.69 \\ \hline
1508420 &       149.83003 &        2.507039 & 24.78 &  4.80 &  0.63 &  -99 &   4.80 &       2 & no & -21.37 \\ \hline
184803 &       149.60240 &        1.580454 & 23.42 &  3.27 &  0.41 &  -99 &   3.27 &       2 & no & -21.39 \\ \hline
316248 &       150.25449 &        1.694429 & 24.49 &  3.40 &  0.42 &  -99 &   3.40 &       2 & no & -20.45 \\ \hline
1314059 &       149.62923 &        2.329145 & 24.59 &  3.40 &  0.39 &  -99 &   3.40 &       2 & no & -20.61 \\ \hline
853930 &       149.56989 &        2.037617 & 24.14 &  3.12 &  0.05 &  -99 &   3.12 &       2 & no & -20.55 \\ \hline
1615630 &       150.55912 &        2.769707 & 24.95 &  3.16 &  3.35 &  -99 &   3.35 &       2 & no & -20.29 \\ \hline
1510107 &       149.84480 &        2.495914 & 24.64 &  3.70 &  0.64 &  -99 &   3.70 &       2 & no & -20.70 \\ \hline
840968 &       149.81004 &        2.124751 & 23.65 &  3.80 &  0.63 &  -99 &   3.80 &       2 & no & -21.93 \\ \hline
418417 &       149.56200 &        1.662382 & 24.05 &  5.30 &  0.96 &  -99 &   5.30 &       2 & no & -22.99 \\ \hline
1596481 &       150.67618 &        2.732485 & 24.85 &  3.40 &  3.12 &  -99 &   3.12 &       2 & no & -19.76 \\ \hline
1045616 &       149.79565 &        2.283589 & 24.80 &  4.10 &  4.21 &  -99 &   4.21 &       2 & no & -20.93 \\ \hline
1385228 &       150.64104 &        2.529159 & 23.51 &  3.30 &  0.62 &  -99 &   3.30 &       2 & no & -21.26 \\ \hline
1552773 &       149.47151 &        2.514272 & 23.66 &  4.14 &  0.21 &  -99 &   4.14 &       2 & no & -21.19 \\ \hline
1893339 &       150.17361 &        2.856197 & 24.49 &  4.26 &  4.22 &  -99 &   4.22 &       2 & no & -20.97 \\ \hline
697377 &       150.59901 &        2.099155 & 23.10 &  3.22 &  0.29 &  -99 &   3.22 &       2 & no & -21.47 \\ \hline
1943641 &       149.78291 &        2.824640 & 24.26 &  5.00 &  5.02 &  -99 &   5.02 &       2 & no & -22.07 \\ \hline
1739756 &       149.60091 &        2.764447 & 24.34 &  5.80 &  5.45 &  -99 &   5.45 &       2 & no & -23.48 \\ \hline
851974 &       149.58417 &        2.046672 & 24.32 &  3.40 &  3.66 &  -99 &   3.66 &       2 & no & -20.98 \\ \hline
1665352 &       150.12608 &        2.767497 & 24.80 &  6.00 &  2.37 &  -99 &   6.00 &       2 & no & -22.25 \\ \hline
1401595 &       150.46024 &        2.577492 & 24.55 &  3.28 &  0.21 &  -99 &   3.28 &       2 & no & -20.43 \\ \hline
784661 &       150.10434 &        2.001693 & 24.21 &  3.30 &  0.27 &  -99 &   3.30 &       2 & no & -20.82 \\ \hline

\end{longtable*}
\end{center}

\end{document}